\documentclass[11pt]{article}

\usepackage[final]{acl}

\usepackage{times}
\usepackage{latexsym}

\usepackage[T1]{fontenc}
\usepackage[utf8]{inputenc}

\usepackage{microtype}

\usepackage{inconsolata}
\usepackage{microtype}
\usepackage{graphicx}
\usepackage{fvextra}
\usepackage{listings}
\usepackage{float}
\definecolor{sjtured}{RGB}{201, 21, 30}
\definecolor{sjtublue}{RGB}{39, 116, 174}

\usepackage{url}
\usepackage{booktabs}       % professional-quality tables
\usepackage{amsfonts}       % blackboard math symbols
\usepackage{nicefrac}       % compact symbols for 1/2, etc.
\usepackage{microtype}      % microtypography
\usepackage{xcolor}         % colors
\usepackage{graphicx}
\usepackage{multirow}
\usepackage{caption}
\usepackage{subcaption}
\usepackage{varwidth}
\usepackage{mwe}
\usepackage{todonotes}
\usepackage{makecell}
\usepackage{amssymb}  % 提供数学符号
\usepackage{pifont}   % 提供 \checkmark 和 \xmark
\usepackage{amsmath}
\usepackage{cleveref}
\usepackage[most]{tcolorbox}
\usepackage{enumitem}
\usepackage{scalerel}

\newcommand{\cmark}{\textcolor{green!60!black}{\ding{51}}} % 绿色对号
\newcommand{\xmark}{\textcolor{red}{\ding{55}}} % 红色叉号

\title{UltraVoice: Scaling Fine-Grained Style-Controlled Speech \\ Conversations for Spoken Dialogue Models}

\author{Wenming Tu$^{1,2}$\footnotemark[1]~, Guanrou Yang$^{1,3}$, Ruiqi Yan$^{1}$, Wenxi Chen$^{1,3}$, Ziyang Ma$^{1,3}$, Yipeng Kang$^{2}$ \\
\textbf{Kai Yu}$^{1}$, \textbf{Xie Chen}$^{1,3}$\footnotemark[2]~, \textbf{Zilong Zheng}$^{2}$\footnotemark[2] \\
$^1$ X-LANCE Lab, Department of Computer Science and Engineering\\ Shanghai Jiao Tong University, Shanghai, China\\
$^2$ State Key Laboratory of General Artificial Intelligence, BIGAI, Beijing, China \\
$^3$ Shanghai Innovation Institute, Shanghai, China \\
\texttt{\small \{tuwenming,chenxie95\}@sjtu.edu.cn, zlzheng@bigai.ai}
}

\begin{document}
\maketitle
\begingroup
\renewcommand{\thefootnote}{\fnsymbol{footnote}}
\footnotetext[1]{Work done during Wenming Tu's internship at BIGAI.}
\footnotetext[2]{Corresponding Author.}
\endgroup
\begin{abstract}
Spoken dialogue models currently lack the ability for fine-grained speech style control, a critical capability for human-like interaction that is often overlooked in favor of purely functional capabilities like reasoning and question answering. To address this limitation, we introduce \textbf{UltraVoice}, the first large-scale spoken dialogue dataset engineered for multiple fine-grained speech style control. Encompassing over 830 hours of spoken dialogues, UltraVoice provides instructions across six key speech stylistic dimensions: emotion, speed, volume, accent, language, and composite styles. Fine-tuning leading models such as SLAM-Omni and VocalNet on UltraVoice significantly enhances their fine-grained speech stylistic controllability without degrading core conversational abilities. Specifically, our fine-tuned models achieve improvements of 29.12-42.33\% in Mean Opinion Score (MOS) and 14.61-40.09 percentage points in Instruction Following Rate (IFR) on multi-dimensional control tasks designed in the UltraVoice.
Moreover, on the URO-Bench benchmark, our fine-tuned models demonstrate substantial gains in core understanding, reasoning, and conversational abilities, with average improvements of +10.84\% on the Basic setting and +7.87\% on the Pro setting. Furthermore, the dataset's utility extends to training controllable Text-to-Speech (TTS) models, underscoring its high quality and broad applicability for expressive speech synthesis. The complete dataset and model checkpoints are available at: \href{https://github.com/bigai-nlco/UltraVoice}{https://github.com/bigai-nlco/UltraVoice}.
\end{abstract}
\begin{figure*}[h!]
    \centering
    \includegraphics[width=\textwidth]{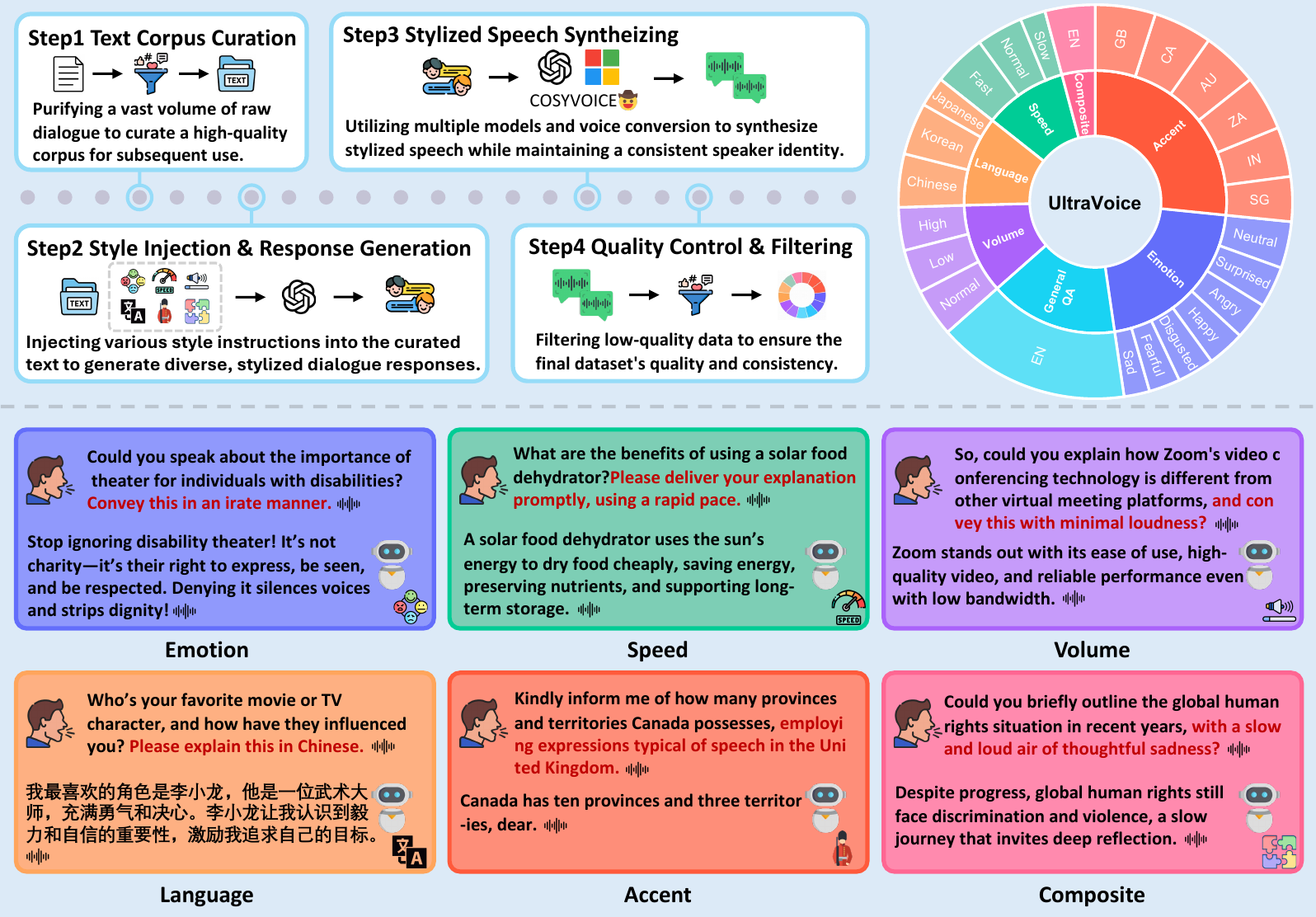}
    \caption{\textbf{Overview of the UltraVoice Dataset Construction and Stylistic Coverage.} The upper left section details the four-step process: text corpus curation, style injection \& response generation, stylized speech synthesis, and quality control \& filtering. The ring chart on the right visualizes the dataset’s control dimensions (inner ring) and their finer control sub-dimensions (outer ring). The lower panel provides examples of six speech style dimensions, including emotion, speed, volume, language, accent, and composite styles (e.g., combinations of speed, volume, and emotion).}
    \label{fig:teaser}
    \vspace{-.1in}
\end{figure*}
\section{Introduction}
The future of human-computer interaction is moving toward more natural, efficient, and expressive communication. With the rise of large language models (LLMs) and their integration with speech technologies, end-to-end spoken dialogue models such as GPT-4o~\citep{GPT-4o}, LL\-aMA-Omni~\citep{fang2024llama,fang2025llama}, and Mini-Omni~\citep{Mini-Omni,Mini-Omni2} have emerged. These models enable real-time, low-latency speech interaction, greatly enhancing user experience. However, most current research has prioritized the functional aspects of conversation (what to say), while the expressive dimension (how to say it) remains largely underdeveloped. Current models can generate fluent responses, but often do so with a neutral or monotonous prosody, lacking the ability to convey nuanced intent, emotion, or personality~\citep{survey-slm,survey-slm2, osum-chat}.

This lack of expressive control is a significant barrier to human-like interaction. For example, imagine a spoken dialogue model generating the response, ``\texttt{\small That's a fantastic idea.}'' Without the ability to precisely control its delivery, the model cannot convey genuine excitement to celebrate a user's suggestion or adopt a playfully sarcastic tone in a creative storytelling scenario. The speech it produces is functionally correct, but expressively sterile. This expressive gap stems from a fundamental flaw in existing training data. The common practice of simply applying Text-to-Speech (TTS) to text-based dialogue datasets fails to inject authentic paralinguistic information. This process results in speech that is linguistically correct but expressively impoverished, lacking the genuine variations in emotion, tone, and prosody that characterize human interaction. Consequently, models are trained on acoustically sterile data, learning what to say, but never learning how to say it with meaningful, human-like expression.

Our primary goal is to significantly enhance the expressiveness of spoken dialogue models by enabling them to modulate their speech style on command. This objective motivates our core research question: \textit{How can we construct a dataset that is sufficiently \textbf{large-scale}, \textbf{diverse}, and \textbf{instruction-rich} to effectively train spoken dialogue models for multi-dimensional, fine-grained speech style control?} We contend that this is achievable, but it requires overcoming the following key challenges.

\textbf{First}, existing spoken dialogue datasets are fundamentally inadequate. Many spoken dialogue datasets, such as InstructS2S~\citep{fang2024llama,fang2025llama} and VoiceAssistant~\citep{Mini-Omni,Mini-Omni2}, are created by simply converting text conversations to speech via TTS. This process yields a mere ``spoken version'' of text, stripped of the authentic, context-driven paralinguistic cues essential for human interaction. This necessitates a new approach beyond simple adaptation. \textbf{Second}, both collecting real data and repurposing existing resources present major obstacles. Acquiring large-scale, real-world spoken dialogues is prohibitively expensive and labor-intensive, while adapting controllable TTS datasets, such as EmoVoice-DB~\citep{EmoVoice}, SpeechCraft~\citep{Speechcraft}, and InstructTTSEval~\citep{instructttseval}, for dialogue is also flawed; forcing them into a conversational format with prompts like, ``Please read this in an excited tone,'' fundamentally degenerates the interactive dialogue task into a non-interactive TTS task. \textbf{Third}, while data synthesis emerges as the most viable path, it presents its own complex hurdles. Its success hinges not only on selecting a sufficiently expressive TTS model \citep{CosyVoice1,CosyVoice2,Spark-TTS,indextts2} to avoid monotonous outputs but, more critically, on a sophisticated generation strategy. This strategy must ensure the authenticity of the generated speech while achieving broad diversity across instructions and control dimensions, ultimately creating data that enables models to learn the nuanced relationship between content (\textit{what} to say) and delivery (\textit{how} to say).

\begin{table*}[t]
\centering
% Make sure you have \usepackage{multirow} and \usepackage{makecell} in your preamble
\resizebox{\textwidth}{!}{%
\begin{tabular}{lccccccccc}
\toprule
\multirow{2.5}{*}{\textbf{Dataset}} & \multirow{2.5}{*}{\textbf{Domain}} & \multirow{2.5}{*}{\centering \makecell{\textbf{General} \\ \textbf{QA}}} & \multicolumn{6}{c}{\textbf{Fine-Grained Control Types}} & \multirow{2.5}{*}{\makecell{\textbf{\#Control} \\ \textbf{Types}}} \\
\cmidrule(lr){4-9}
& & & \textbf{Emotion} & \textbf{Speed} & \textbf{Volume} & \textbf{Language} & \textbf{Accent} & \textbf{Composite} & \\
\midrule
SpeechCraft~(\citeyear{Speechcraft}) & \multirow{3}{*}{\makecell{Controllable \\ TTS}} & \xmark & \xmark & \xmark & \xmark & \xmark & \xmark & \cmark & 1 \\
EmoVoice-DB~(\citeyear{EmoVoice}) & & \xmark & \cmark & \xmark & \xmark & \xmark & \xmark & \xmark & 1 \\
LAION's Got Talent\footnotemark & & \xmark & \cmark & \xmark & \xmark & \xmark & \xmark & \xmark & 1 \\
\midrule
InstructS2S~(\citeyear{fang2024llama}) & \multirow{3}{*}{\makecell{Spoken \\ Dialogue}} & \cmark & \xmark & \xmark & \xmark & \xmark & \xmark & \xmark & 0 \\
VoiceAssistant~(\citeyear{Mini-Omni}) & & \cmark & \xmark & \xmark & \xmark & \xmark & \xmark & \xmark & 0 \\
\textbf{UltraVoice (Ours)} & & \cmark & \cmark & \cmark & \cmark & \cmark & \cmark & \cmark & 6 \\
\bottomrule
\end{tabular}
}
\caption{Comparison of Speech Datasets in Terms of Fine-Grained Speech Style Control}
\label{tab:speech_datasets}
\end{table*}

To address the above challenges, this work makes three core contributions. \textbf{Firstly}, we introduce \textbf{UltraVoice}, the first large-scale spoken dialogue dataset engineered for multiple fine-grained speech style control, filling a key gap in the field. It supports fine-grained control across six stylistic dimensions, including emotion, volume, speed, accent, language, and composite styles, providing a solid basis for training and evaluating expressive spoken dialogue models. \textbf{Secondly}, we conduct comprehensive supervised fine-tuning (SFT) on mainstream spoken dialogue models on it, and we observe consistent gains in expressive style rendering and general conversational competence. \textbf{Thirdly}, we demonstrate the dataset’s generalizability beyond dialogue modeling by fine-tuning a pre-trained TTS model, which enables multidimensional controllable speech synthesis across diverse styles and highlights the dataset’s versatility and reliability for downstream speech generation.

\section{Related Work}
\footnotetext{https://huggingface.co/datasets/laion/laions\_got\_talent}
\subsection{End-to-End Spoken Dialogue Models}

Early end-to-end spoken dialogue models sought to integrate automatic speech recognition (ASR), text-based dialogue modeling, and text-to-speech synthesis (TTS) within a unified architecture to reduce inference latency~\citep{audiogpt,funaudiollm}. Pioneering models such as Mini-Omni~\citep{Mini-Omni,Mini-Omni2} and Moshi~\citep{Moshi} adopted shared decoders that jointly generate text and audio tokens, while later models, including LLaMA-Omni~\citep{fang2024llama,fang2025llama} and Freeze-Omni~\citep{Freeze-Omni} employed modular multimodal pipelines with dedicated speech encoders and decoders built around a pre-trained LLM. Despite these architectural advances, style controllability remains a significant weakness. Since expressiveness is learned implicitly from training data, these models tend to produce homogeneous speaking styles and lack explicit control over paralinguistic features such as emotion and speed. This deficiency severely limits their use in personalized or emotionally expressive dialogue settings~\citep{wavchat}.

Recent models~\citep{Qwen2.5-Omni,step-audio} have begun to address these limitations. For instance, SLAM-Omni~\citep{SLAM-OMNI} introduces zero-shot timbre control, enabling real-time dialogue with dynamic speaker voices specified via audio prompts.  On the efficiency front, VocalNet~\citep{VocalNet} enhances both generation speed and quality through multi-token prediction (MTP), producing multiple audio tokens per decoding step rather than one at a time. Nonetheless, none of the existing end-to-end spoken dialogue models provides explicit fine-grained speech style controls such as direct modulation of emotion, accent, or speed. In summary, while end-to-end dialogue models have made substantial progress in generating natural and low-latency speech, the ability to explicitly manipulate stylistic attributes remains entirely unaddressed in current approaches.

\subsection{Spoken Dialogue and Controllable TTS Dataset}

For general-purpose spoken dialogue tasks, existing datasets mainly prioritize functionality over expressiveness. Well-known corpora such as InstructS2S~\citep{fang2024llama} and VoiceAssistant~\citep{Mini-Omni} have been widely adopted to train models, including LLaMA-Omni and Mini-Omni, supporting task-oriented interactions, voice assistants, and related applications. These datasets typically contain hundreds of thousands of spoken dialogue pairs and enable direct speech interaction. Despite their scale and dialogue focus, they lack explicit fine-grained speech style annotations, such as speed, volume, or emotion. As a result, the generated speech is often homogeneous and lacks the fine-grained control required for emotionally rich or personalized interactions.

Controllable TTS datasets, such as Speech-Craft~\citep{Speechcraft} for description-based synthesis and EmoVoice-DB~\citep{EmoVoice} for emotional control, are designed to produce speech with specific styles. The recent success of state-of-the-art models~\citep{controllablettssurvey} trained on such data, including CosyVoice~\citep{CosyVoice1,CosyVoice2,Cosyvoice3} and the audio model of GPT-4o-audio-preview~\citep{GPT-4o}, highlights the significant progress in fine-grained stylistic generation. However, a fundamental limitation of these datasets persists: they are overwhelmingly designed for non-interactive synthesis. Because these corpora lack the bidirectional dialogue structure and turn-taking context inherent to conversation, they are ultimately unsuitable for training end-to-end spoken dialogue models.

To address the lack of explicit speech style control instructions in spoken dialogue datasets and the non-interactive limitation of TTS corpora, we introduce \textbf{UltraVoice}. As summarized in~\Cref{tab:speech_datasets}, UltraVoice covers six key stylistic dimensions: emotion, speed, volume, language, accent, and composite styles (e.g., combinations of speed, volume, and emotion). It also maintains full dialogue context along with instruction and response structure. This dataset fills a critical gap by supporting both general speech interaction and fine-grained speech style control, providing a unified and high-quality dataset for training and evaluating style-controllable end-to-end spoken dialogue models.

\section{The UltraVoice Dataset}\label{sec:dataset}

To facilitate a deeper understanding of our dataset construction pipeline, this section offers a comprehensive overview of the four key steps involved in building UltraVoice, as illustrated in~\Cref{fig:teaser}. We have designed a bottom-up, fine-grained data generation pipeline that spans text preparation, style instruction injection, speech synthesis, and data filtering. This pipeline integrates everyday conversational texts with a wide range of speech style control types, ensuring high consistency and diversity in content, vocal style, and audio quality. The following subsections will elaborate on the core tasks and implementation details of each step.

\textbf{Step 1: Text Corpus Curation.}
To construct the UltraVoice dataset, we curated the foundational text corpus using UltraChat~\citep{UltraChat}, a widely adopted English dialogue dataset frequently used for spoken dialogue synthesis in models such as LLaMA-Omni~\citep{fang2024llama,fang2025llama}, Mini-Omni~\citep{Mini-Omni,Mini-Omni2}, and SLAM-Omni~\citep{SLAM-OMNI}. We extracted dialogues primarily from the \textit{Question About the World} and \textit{Creation and Generation} categories due to their conciseness and independence from external references. To ensure high-quality input for downstream synthesis, we applied strict filtering rules to remove dialogues containing URLs, academic citations, or lengthy quoted texts. After filtering, we obtained approximately 200,000 clean and natural question-answer pairs, which served as the base for style-controlled speech generation.

\textbf{Step 2: Style Injection \& Response Generation.}
To enable fine-grained control over speaking styles, we predefined six stylistic dimensions: speed, volume, emotion, accent, language, and composite styles. For each dimension, we used GPT-4o~\citep{GPT-4o} to generate diverse and natural style prompts (see \Cref{app:prompts} for all prompt templates), leveraging semantically similar expressions (e.g., “respond in a joyful tone” vs. “reply with a cheerful voice”). Based on these prompts, GPT-4o was further invoked to generate stylized textual responses, ensuring alignment in semantics and tone. Additionally, we applied several practical adjustments to improve downstream TTS following~\citet{fang2024llama}. For example, we expanded numbers into spoken forms (e.g., “123” → “one two three”) and rephrased code-related queries to encourage natural spoken language responses. These refinements ensured better speech synthesis fluency and user interaction quality.

\textbf{Step 3: Stylized Speech Synthesizing.}
In this step, we performed speech synthesis for each instruction-response speech pair to simulate realistic conversations with fine-grained style control. For instruction speech, we randomly sampled speaker timbres from the seedtts\_testset\_en\footnote{https://github.com/BytedanceSpeech/seed-tts-eval}~\citep{Seed-tts} corpus. This corpus features diverse speakers and real-world background noise, allowing the instruction audio to better reflect realistic user conditions. In contrast, response speech was synthesized using a single fixed timbre to ensure consistency across all stylized outputs. We selected the TTS model for each style control dimension as detailed in~\Cref{tab:tts_control}. Most responses were synthesized using the GPT-4o-audio-preview~\citep{GPT-4o} model due to its expressiveness and high fidelity. For accent-specific responses, we used Edge TTS\footnote{https://github.com/rany2/edge-tts}, which lacks support for custom speaker timbres. To address this, we applied voice conversion (VC) via CosyVoice-300M~\citep{CosyVoice1} to align the output with the designated fixed voice.
To ensure data balance, we further sampled 40,000 generic QA pairs without style instructions from the VoiceAssistant400k\footnote{https://huggingface.co/datasets/gpt-omni/VoiceAssistant-400K} corpus. After removing templated phrases (e.g., “I’m mini omni”), we resynthesized them using CosyVoice-300M.
\begin{table}[h]
  \centering
  \small % 减小字号以适应单栏
  \resizebox{\linewidth}{!}{ % 如果还是太宽，取消注释此行或保留
    \begin{tabular}{lcc}
      \toprule
      \makecell{\textbf{Control} \\ \textbf{Dimension}} & \makecell{\textbf{Instruction} \\ \textbf{TTS Model}} & \makecell{\textbf{Response} \\ \textbf{TTS Model}} \\
      \midrule
      Accent & \multirow{7}{*}{CosyVoice 300M} & Edge TTS + VC \\
      Composite & & GPT-4o-audio-preview \\
      Emotion & & GPT-4o-audio-preview \\
      Language & & CosyVoice 300M \\
      Speed & & GPT-4o-audio-preview \\
      Volume & & GPT-4o-audio-preview \\
      General QA & & CosyVoice 300M \\
      \bottomrule
    \end{tabular}
  }
  \caption{TTS model selections for different dimensions.}
  \label{tab:tts_control}
\end{table}

% \begin{figure*}[htbp!] 
%   \centering
%   % 左侧两个子图
%   \begin{minipage}[b]{0.58\textwidth}
%     \centering
%     \begin{subfigure}[b]{0.49\textwidth}
%       \centering
%       \includegraphics[width=\linewidth]{figs/duration_distribution.pdf}
%       \caption{Duration Distribution}
%       \label{fig:sub1}
%     \end{subfigure}
%     \hfill
%     \begin{subfigure}[b]{0.49\textwidth}
%       \centering
%       \includegraphics[width=\linewidth]{figs/word_cnt_distribution.pdf}
%       \caption{Word Count Distribution}
%       \label{fig:sub2}
%     \end{subfigure}
%     \caption{Distributions of Duration and Number of Words.}
%     \label{fig:duration_word_dist}
%   \end{minipage}
%   \hfill
%   % 右侧表格
%   \begin{minipage}[b]{0.39\textwidth}
%     \centering
%     \resizebox{\linewidth}{!}{
%       \begin{tabular}{lcc}
%         \toprule
%         \makecell{\textbf{Control} \\ \textbf{Dimension}} & \makecell{\textbf{Instruction} \\ \textbf{TTS Model}} & \makecell{\textbf{Response} \\ \textbf{TTS Model}} \\
%         \midrule
%         Accent & \multirow{7}{*}{CosyVoice 300M} & Edge TTS + VC \\
%         Composite & & GPT-4o-audio-preview \\
%         Emotion & & GPT-4o-audio-preview \\
%         Language & & CosyVoice 300M \\
%         Speed & & GPT-4o-audio-preview \\
%         Volume & & GPT-4o-audio-preview \\
%         General QA & & CosyVoice 300M \\
%         \bottomrule
%       \end{tabular}
%     }
%     \captionof{table}{TTS model selections for different control dimensions.}
%     \label{tab:tts_control}
%   \end{minipage}
% \end{figure*}

\textbf{Step 4: Quality Control \& Filtering.}
To ensure the overall quality and balanced stylistic coverage of the dataset, we applied an automated filtering process to all synthesized spoken dialogue samples. Specifically, we utilized the Whisper-large-v3~\citep{Whisper} to perform automatic speech recognition (ASR) on each instruction and response audio sample, and computed the character error rate (CER) based on the transcriptions. We applied a unified data filtering criterion to both instruction and response audio: only retaining samples with a CER below 20\% and duration under 30 seconds. This filtering pipeline effectively removed samples with high ASR error or abnormal length, significantly improving the dataset’s consistency and usability.

\subsection{Characteristics and Statistics}
\begin{table*}[h!]
  \centering
  \small

  \setlength{\tabcolsep}{5.5pt}
  \resizebox{1\textwidth}{!}{%
  \begin{tabular}{@{}l p{0.53\linewidth} r r c c @{}}
    \toprule
    \textbf{Dimension} & \textbf{Fine-grained Control Dimensions} & \textbf{\#Cnt.} & \textbf{Dur.(h)} & \textbf{CER} & \textbf{UTMOS} \\
    \midrule
    Emotion & Neutral, Happy, Sad, Angry, Surprised, Fearful, Disgusted & 21,209 & 182.53 & 6.17 & 3.98\\
    Volume & Low Volume, High Volume, Normal Volume & 11,154 & 91.37 & 5.29 & 3.80\\
    Speed  & Slow Speed, Fast Speed, Normal Speed & 10,334 & 85.28 & 4.84 & 4.05\\
    Accent & AU, CA, GB, IN, SG, ZA & 26,839 & 253.31 & 6.69 & 4.08\\
    Language & Chinese, Korean, Japanese & 11,153 & 93.84 & 6.83 & 3.95\\
    Composite & Combinations of speed, volume, and emotion & 4,143 & 33.47 & 5.02 & 3.97\\
    General QA & English general question answering & 15,938 & 93.12 & 6.69 & 4.15\\
    \midrule
    \textbf{Overall} & & \textbf{100,770} & \textbf{832.92} & \textbf{5.93} & \textbf{4.00}\\
    \bottomrule
  \end{tabular}
  }
  \caption{\textbf{Detailed statistics of UltraVoice across different control dimensions.} \textbf{\#Cnt.} denotes the number of samples, \textbf{Dur.} is the total duration in \textbf{hours}, \textbf{CER} is the average character error rate, and \textbf{UTMOS} represents the averaged naturalness score. \textbf{AU}, \textbf{CA}, \textbf{GB}, \textbf{IN}, \textbf{SG}, and \textbf{ZA} correspond to accents from Australia, Canada, United Kingdom, India, Singapore, and South Africa, respectively.}
\label{tab:ctrl-attributes}
\end{table*}

As summarized in Table~\ref{tab:ctrl-attributes}, the UltraVoice dataset comprises 100,770 spoken dialogue samples, among which 84,832 are explicitly conditioned on six major control dimensions: emotion, volume, speed, accent, language, and composite styles description. The remaining 15,938 pairs are general English QA samples without style prompts, added to improve balance and generalization. The dataset includes 21,209 emotion-controlled samples across seven categories (\textit{neutral, happy, sad, angry, surprised, fearful, disgusted}), 11,154 for volume (\textit{low, normal, high}), and 10,334 for speed (\textit{slow, normal, fast}). Accent control covers six English variants (\textit{AU, CA, GB, IN, SG, ZA}) totaling 26,839 samples. Language-switching samples span Chinese, Japanese, and Korean, with 11,153 entries. The composite style dimension includes 4,143 samples for multidimensional control (\textit{e.g., respond with a slow and loud air of thoughtful sadness}).

% \begin{figure}[h] % 使用 [t] 或 [htbp] 放在单栏顶部
%   \centering
%   \begin{subfigure}[b]{0.49\linewidth}
%     \centering
%     \includegraphics[width=\linewidth]{figs/duration_distribution.pdf}
%     \caption{Duration}
%     \label{fig:sub1}
%   \end{subfigure}
%   \hfill
%   \begin{subfigure}[b]{0.49\linewidth}
%     \centering
%     \includegraphics[width=\linewidth]{figs/word_cnt_distribution.pdf}
%     \caption{Word Count}
%     \label{fig:sub2}
%   \end{subfigure}
%   \caption{Duration and Word Count Distributions}
%   \label{fig:duration_word_dist}
% \end{figure}

\begin{figure}[h] % 使用 [t] 或 [htbp] 放在单栏顶部
  \centering
  \begin{subfigure}[b]{0.49\linewidth}
    \centering
    \includegraphics[width=\linewidth]{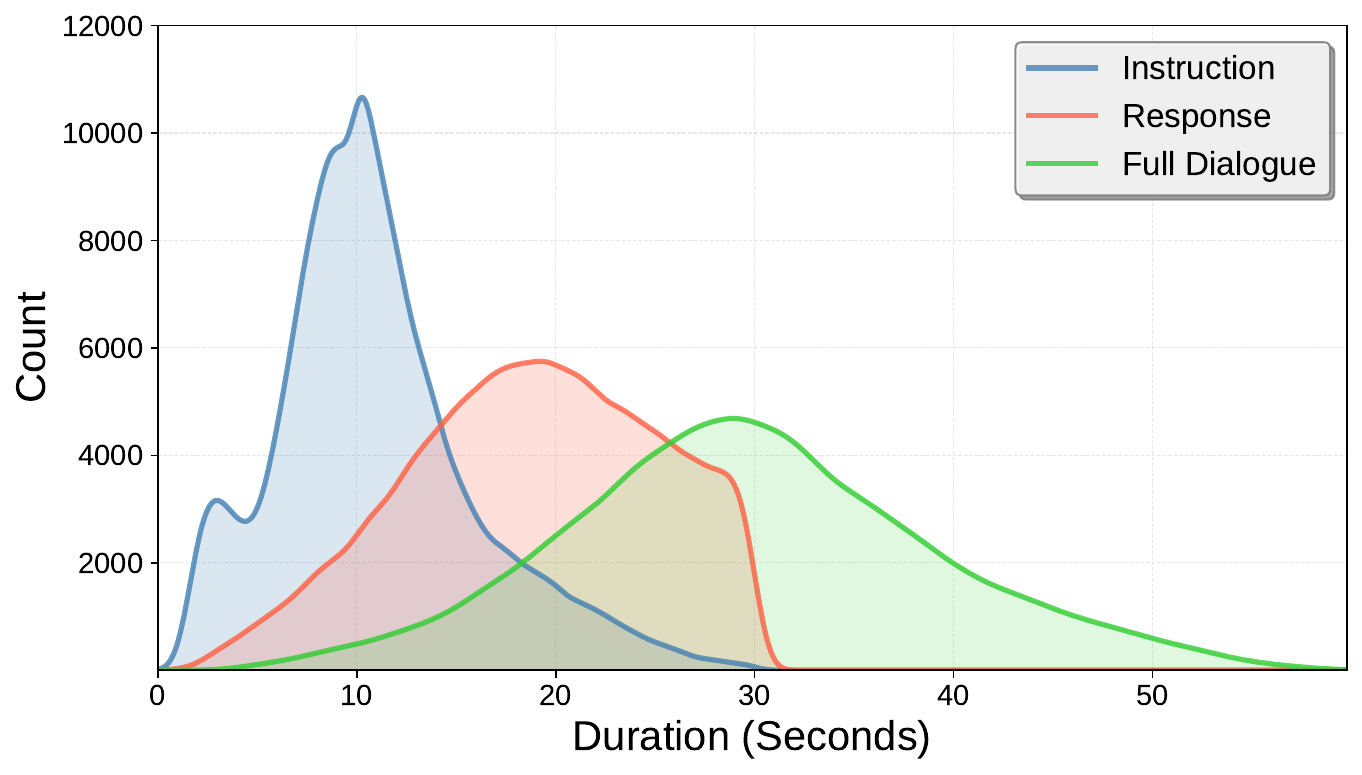}
    \caption{Duration}
    \label{fig:sub1}
  \end{subfigure}
  \hfill
  \begin{subfigure}[b]{0.49\linewidth}
    \centering
    \includegraphics[width=\linewidth]{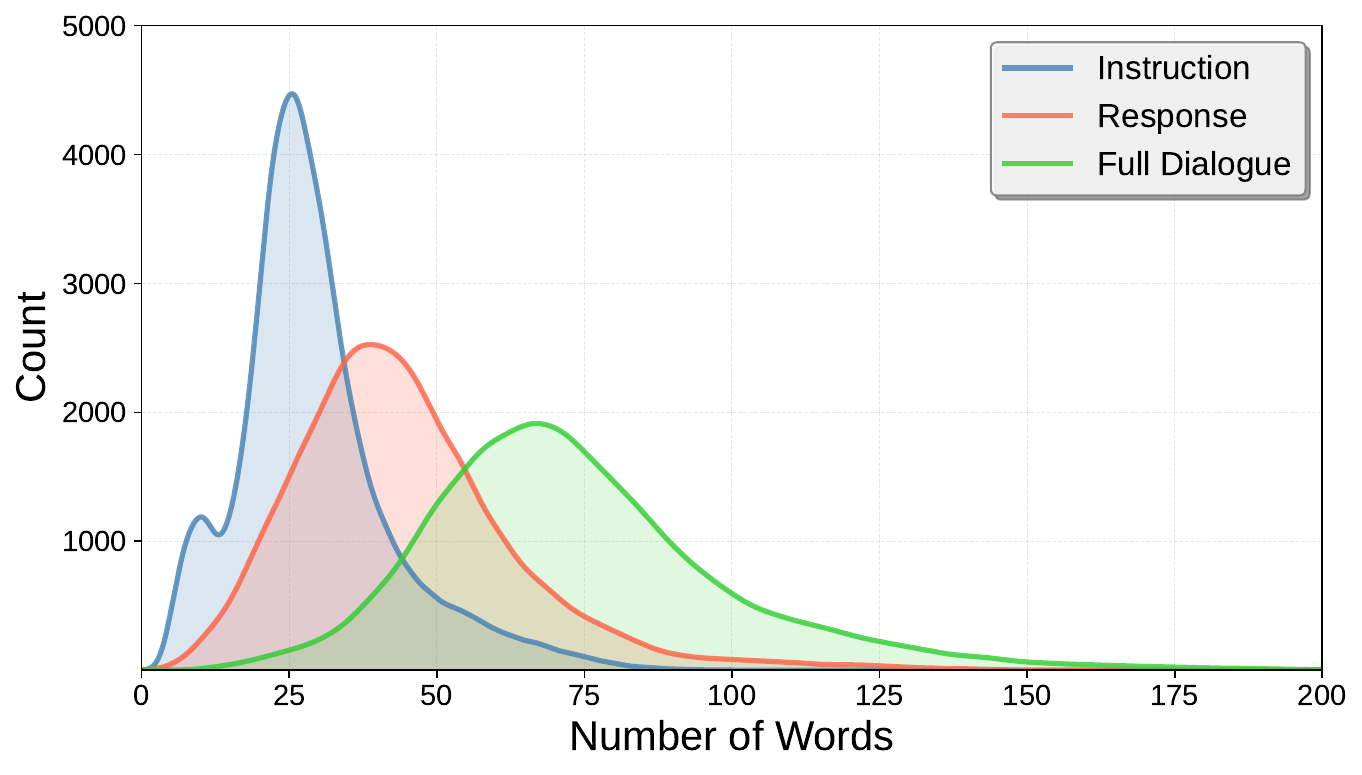}
    \caption{Word Count}
    \label{fig:sub2}
  \end{subfigure}
  \caption{Duration and Word Count Distributions}
  \label{fig:duration_word_dist}
\end{figure}
In total, the dataset covers over 830 hours of speech, with duration distribution shown in ~\Cref{fig:duration_word_dist}. Alongside duration, we also report word count distributions to assess utterance complexity and length variation. The structured control space and balanced temporal characteristics make UltraVoice a valuable resource for training and evaluating stylistically controllable spoken dialogue systems. More detailed statistics are  in \Cref{app:detailed_datasest_statistics}.

\subsection{Quality Assessment}
% wer, utmos, duration, tsne{emotion, accent}, objective{speed volume} 
% quality assessment就介绍CER、Duration、UTMOS、Word Count这些信息

To ensure the quality and consistency of the spoken dialogue data, we applied strict filtering criteria, retaining only samples with a duration under 30 seconds and a CER below 20\%. This filtering removes samples with unreliable ASR outputs or abnormal content. As reported in Table~\ref{tab:ctrl-attributes}, the final dataset achieves an average dialogue length of 29.35 seconds, a mean CER of 5.93\%, and an overall UTMOS~\citep{utmos} score of 4.00, indicating high naturalness and stylistic fidelity of the generated speech.

Beyond quantitative metrics (average duration 29.35s, mean CER 5.93\%, UTMOS 4.00), we provide visual analyses to further validate data quality. As shown in~\Cref{fig:quality_assessment}, six visualization types assess the effectiveness and clarity of our style control design. Emotion and accent are visualized using t-SNE plots from classification model features, showing clear category separability. Speed and volume are illustrated via word-per-minute (WPM) and root-mean-square (RMS) distributions, confirming consistent prosodic control. Language and composite styles are represented with word clouds, showcasing lexical diversity and expressive richness. These results demonstrate the robustness and interpretability of UltraVoice’s stylistic control.

% \newpage
\section{Experiment}\label{sec:experiment}
We systematically evaluate the performance of the spoken dialogue models fine-tuned on UltraVoice. Firstly, we verify the model’s ability to control multi-dimensional speech styles on the UltraVoice internal test set after SFT. Next, we examine the model’s generalization capability on the URO-Bench~\citep{URO-Bench}. Finally, we further validate the quality of our fine-grained controlled response speech by successfully training a controllable TTS model following the pipeline of EmoVoice~\citep{EmoVoice} on a dataset constructed from the UltraVoice.

\subsection{Experiment Setup}

\begin{figure*}[t]
    \centering
    \includegraphics[width=\linewidth]{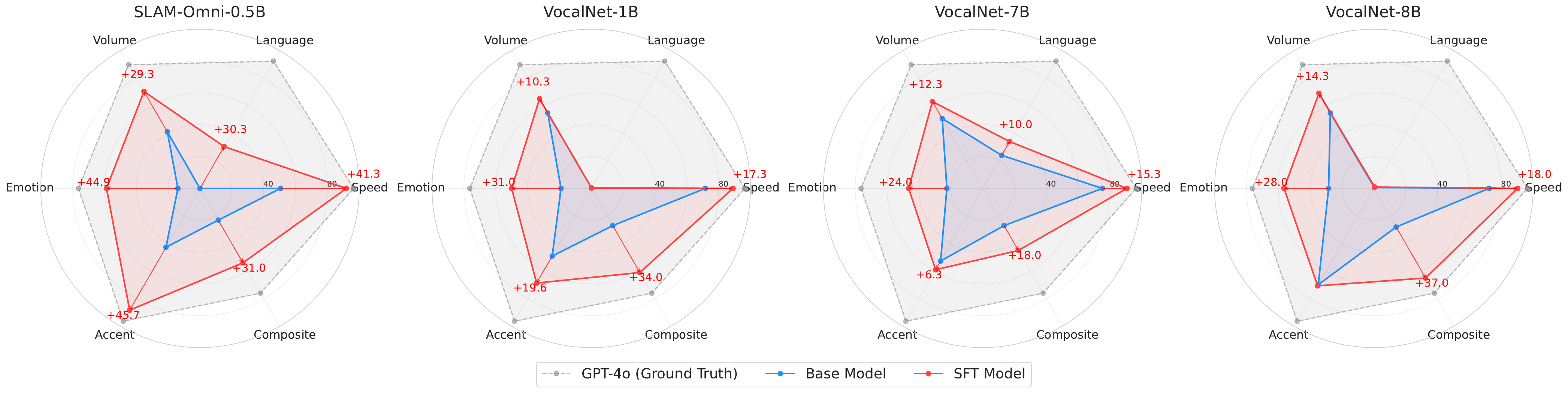}
    \caption{\textbf{IFR (\%) results across six fine-grained speech style control dimensions for each model.} Each radar chart contrasts the base model (\textcolor[HTML]{4D8BFF}{Blue}) and its SFT variant (\textcolor[HTML]{FF4B4B}{Red}), with GPT-4o (\textcolor[HTML]{808080}{Gray}) used as a reference.}
    \label{fig:instr_acc}
\end{figure*}

\textbf{Settings.}
We evaluate four spoken dialogue models from the SLAM-Omni~\citep{SLAM-OMNI} and VocalNet~\citep{VocalNet} series, covering different model sizes and LLaMA/Qwen backbones. We applied SFT to these models to analyze their performance on speech style control.  The detailed configurations of the spoken dialogue models (~\Cref{tab:model-settings}) and the training configurations for SFT (~\Cref{tab:slamomni_sft_config,tab:vocalnet_sft_config,tab:ultravoice_sft_config}) are provided in~\Cref{app:sft_config}.

\textbf{Evaluation and metrics.}
We construct the UltraVoice test set by randomly sampling 100 examples from each fine-grained dimension across the six major control dimensions, resulting in 2,300 samples. The test set has no overlap with the training data. To assess whether UltraVoice SFT affects general spoken dialogue abilities, we utilized the URO-Bench~\citep{URO-Bench}, which evaluates models across three dimensions: Oral Conversation, Understanding, and Reasoning. It allows us to analyze whether core dialogue competencies are preserved and whether expressive performance improves after fine-tuning.

\underline{\textit{Audio-Language Model (ALM) based Metric.}}
Following the evaluation paradigm similar to methodologies proposed by \citet{URO-Bench,EmoVoice}, we employed Gemini-2.5-Flash~\citep{Gemini2.5} as our automatic evaluator to generate \textbf{Mean Opinion Scores (MOS)} and compute the \textbf{instruction-following rate (IFR)} for each control dimension. This choice is motivated by findings that advanced ALMs show high consistency with human ju\-dg\-me\-nts by~\citet{ALM-as-judges}. 
% \rev{We further validate this evaluation protocol through a human subjective evaluation.}
% \rev{We further validate this evaluation protocol through a human subjective evaluation, which confirms the consistency of the ALM-based MOS and IFR trends.}
We further validate this evaluation protocol through a human subjective evaluation, which confirms the consistency of the ALM-based MOS and IFR trends.
The human study protocol and complete results are reported in~\Cref{app:human_eval}, while details of the evaluation prompts are provided in~\Cref{app:ifr_eval,app:mos_eval}.
% Details of prom\-pts are ava\-il\-able in~\Cref{app:ifr_eval,app:mos_eval}.

\underline{\textit{Objective Numerical Metric.}}
Content consistency was measured by the \textbf{Word Error Rate (WER)}, using transcriptions from the Whisper-lar\-ge-v3 model~\citep{Whisper}. For emotional expressiveness, we adopted metrics from~\citet{EmoVoice}, leveraging emo\-tion2vec~\citep{emotion2vec} to compute both \textbf{Emotion Similarity} (cosine similarity between embeddings) and \textbf{Recall Rate} (from emotion classification). Finally, the overall naturalness and perceptual audio quality were evaluated using the \textbf{UTMOS} score~\citep{utmos}.

\subsection{Performance on Fine-grained Speech Style Control}

\textbf{Enhancements of Multi-Dim\-en\-sional In\-struc\-tion-Fol\-low\-ing Capabilities. }As shown in \Cref{fig:instr_acc} and detailed in~\Cref{tab:instruction_following_rate} in the~\Cref{app:detailed_performance_comparison}, fine-tuning with UltraVoice significantly boosts the spoken dialogue models' instruction-following capability for fine-grained speech style control, with IFR gains ranging from 14.61 to 40.09 points. This improvement is particularly pronounced for smaller models with weaker baseline performance. For instance, the IFR of SLAM-Omni-0.5B surged from 28.30\% to 68.39\%, while VocalNet-1B's score increased from 36.28\% to 55.91\%. These results demonstrate that even resource-constrained models can achieve substantial gains in responsiveness to control instructions through UltraVoice. Concurrently, larger models such as VocalNet-7B and 8B also exhibited consistent improvements, indicating an enhanced ability to precisely control various dimensions of fine-grained speech styles.

\textbf{Enhancements in the Subjective Naturalness of Fine-Grained Speech Styles Control. }As shown in \Cref{tab:mos_control_eval}, all models exhibit significant improvements in MOS after being fine-tuned with UltraVoice. The relative gains range from 29.12\% to 42.33\%, with the Emotion and Volume dimensions showing particularly remarkable improvements. For instance, the overall MOS for VocalNet-7B increased from 2.73 to 3.59, while VocalNet-8B's score rose from 2.85 to 3.68. These results indicate that our fine-tuning process enhances the models' ability to render the specified styles with high naturalness, without sacrificing audio quality.

\textbf{Cross-Metric Consistency and Limitations. }Overall, MOS and IFR trends are strongly aligned, suggesting that stronger instruction adherence typically yields more natural speech.
% \rev{This trend is further supported by our human subjective evaluation in~\Cref{app:human_eval}, where human judgments show consistent gains with the ALM-based results.}
This trend is further supported by our human subjective evaluation in~\Cref{app:human_eval}, where human judgments show consistent gains with the ALM-based results.
However, the Language control is a notable exception. LLaMA-based models (e.g., VocalNet-1B and 8B) exhibit a slight MOS decline and stagnant IFR, while Qwen-based models (e.g., SLAM-Omni-0.5B and VocalNet-7B) achieve clear gains. This discrepancy likely stems from different multilingual pretraining exposure. Since Language control requires full responses in another language, it poses unique generalization challenges. The current fine-tuning data lacks sufficient multilingual diversity and volume. Future work should explore targeted multilingual instruction augmentation.

\begin{table}[t] % 去掉星号，改为单栏环境
\centering
\small
\addtolength{\tabcolsep}{-4.8pt} % 关键：收紧列间距以适应单栏
\resizebox{\columnwidth}{!}{ % 缩放到单栏宽度
\begin{tabular}{lccccccc}
\toprule
\textbf{Model} & \textbf{Speed} & \textbf{Lang.} & \textbf{Vol.} & \textbf{Emo.} & \textbf{Acc.} & \textbf{Comp.} & \textbf{Avg.} \\
\midrule
\textbf{GPT-4o (GT)} & 4.82 & 4.46 & 4.60 & 4.57 & 4.68 & 4.46 & 4.60 \\
\cmidrule(lr){1-8}
SLAM-Omni-0.5B Base      & 2.58 & 1.13 & 2.47 & 2.17 & 2.23 & 2.33 & 2.15 \\
SLAM-Omni-0.5B SFT       & 3.61 & 1.18 & 3.47 & 3.32 & 3.42 & 3.37 & 3.06 \\
$\Delta$ (\%)       & \textcolor{teal}{+39.9} & \textcolor{teal}{+4.4} & \textcolor{teal}{+40.5} & \textcolor{teal}{+53.0} & \textcolor{teal}{+53.4} & \textcolor{teal}{+44.6} & \textcolor{teal}{+42.3} \\
\cmidrule(lr){1-8}
VocalNet-1B Base  & 3.45 & 1.18 & 3.10 & 2.42 & 2.74 & 2.83 & 2.62 \\
VocalNet-1B SFT   & 4.28 & 1.01 & 3.98 & 3.73 & 3.77 & 3.95 & 3.45 \\
$\Delta$ (\%)       & \textcolor{teal}{+24.1} & \textcolor{red}{-14.4} & \textcolor{teal}{+28.4} & \textcolor{teal}{+54.1} & \textcolor{teal}{+37.6} & \textcolor{teal}{+39.6} & \textcolor{teal}{+31.7} \\
\cmidrule(lr){1-8}
VocalNet-7B Base  & 3.75 & 1.64 & 2.80 & 2.42 & 3.13 & 2.61 & 2.73 \\
VocalNet-7B SFT   & 4.25 & 2.19 & 3.95 & 3.79 & 3.88 & 3.51 & 3.59 \\
$\Delta$ (\%)       & \textcolor{teal}{+13.3} & \textcolor{teal}{+33.5} & \textcolor{teal}{+41.1} & \textcolor{teal}{+56.6} & \textcolor{teal}{+24.0} & \textcolor{teal}{+34.5} & \textcolor{teal}{+31.5} \\
\cmidrule(lr){1-8}
VocalNet-8B Base  & 3.57 & 1.17 & 3.12 & 2.90 & 3.47 & 2.86 & 2.85 \\
VocalNet-8B SFT   & 4.52 & 1.02 & 4.21 & 4.10 & 4.19 & 4.07 & 3.68 \\
$\Delta$ (\%)       & \textcolor{teal}{+26.6} & \textcolor{red}{-12.8} & \textcolor{teal}{+34.9} & \textcolor{teal}{+41.4} & \textcolor{teal}{+20.8} & \textcolor{teal}{+42.3} & \textcolor{teal}{+29.1} \\
\bottomrule
\end{tabular}
}
\caption{\textbf{MOS results across six fine-grained speech style dimensions for each model.} The third row of each group shows the relative gain (\%) from SFT.}
\label{tab:mos_control_eval}
\end{table}

\subsection{General Conversational Ability}

\begin{table*}[h]
    \centering
    \small
    \resizebox{0.93\textwidth}{!}{%
    \begin{tabular}{l cccc cccc}
    \toprule
    \multirow{2}{*}{\textbf{Models}} &
    \multicolumn{4}{c}{\textbf{Basic}} &
    \multicolumn{4}{c}{\textbf{Pro}} \\
    \cmidrule(lr){2-5} \cmidrule(lr){6-9}
    & \textbf{ Und. ↑} & \textbf{Reasoning ↑} & \textbf{Conv. ↑} & \textbf{Avg. ↑} &
      \textbf{Und. ↑} & \textbf{Reasoning ↑} & \textbf{Conv. ↑} & \textbf{Avg. ↑} \\
    \midrule
    SLAM-Omni-0.5B Base & 26.60 & 23.36 & 47.54 & 32.50 & 25.79 & 24.72 & 29.93 & 26.81 \\
    SLAM-Omni-0.5B SFT  & 31.51 & 24.58 & 50.14 & 35.41 & 26.30 & 20.07 & 35.57 & 27.31 \\
    $\Delta$ (\%)       & \textcolor{teal}{+18.46} & \textcolor{teal}{+5.22} & \textcolor{teal}{+5.47} & \textcolor{teal}{+8.95} & \textcolor{teal}{+1.98} & \textcolor{red}{-18.81} & \textcolor{teal}{+18.84} & \textcolor{teal}{+1.87} \\
    \midrule
    VocalNet-1B Base & 58.34 & 41.69 & 66.84 & 55.62 & 34.88 & 46.86 & 38.96 & 40.23 \\
    VocalNet-1B SFT  & 70.41 & 45.19 & 70.81 & 62.14 & 36.06 & 51.42 & 41.08 & 42.85 \\
    $\Delta$ (\%)        & \textcolor{teal}{+20.69} & \textcolor{teal}{+8.40} & \textcolor{teal}{+5.94} & \textcolor{teal}{+11.73} & \textcolor{teal}{+3.38} & \textcolor{teal}{+9.73} & \textcolor{teal}{+5.44} & \textcolor{teal}{+6.51} \\
    \midrule
    VocalNet-7B Base & 81.50 & 64.08 & 78.41 & 74.66 & 37.90 & 58.87 & 45.24 & 47.34 \\
    VocalNet-7B SFT  & \textbf{88.71} & \textbf{71.85} & \textbf{84.12} & \textbf{81.56} & \textbf{46.39} & \textbf{64.52} & 47.20 & 52.70 \\
    $\Delta$ (\%)        & \textcolor{teal}{+8.85} & \textcolor{teal}{+12.13} & \textcolor{teal}{+7.28} & \textcolor{teal}{+9.24} & \textcolor{teal}{+22.40} & \textcolor{teal}{+9.60} & \textcolor{teal}{+4.33} & \textcolor{teal}{+11.32} \\
    \midrule
    VocalNet-8B Base & 65.52 & 53.56 & 75.57 & 64.88 & 37.96 & 53.32 & 42.43 & 44.57 \\
    VocalNet-8B SFT  & 72.37 & 61.52 & 80.87 & 71.59 & 40.57 & 62.07 & 48.50 & 50.38 \\
    $\Delta$ (\%)        & \textcolor{teal}{+10.45} & \textcolor{teal}{+14.86} & \textcolor{teal}{+7.01} & \textcolor{teal}{+10.34} & \textcolor{teal}{+6.88} & \textcolor{teal}{+16.41} & \textcolor{teal}{+14.31} & \textcolor{teal}{+13.04} \\
    \midrule\midrule
    Qwen2.5-Omni-7B  & 66.29 & 69.62 & 76.16 & 70.69 & 44.51 & 63.88 & 49.41 & 52.60 \\
    LLaMA-Omni-8B    & 47.45 & 36.03 & 64.98 & 49.49 & 28.85 & 47.62 & 34.47 & 36.98 \\
    GLM4-Voice-9B    & 82.16 & 55.46 & 74.20 & 70.61 & 45.14 & 61.28 & \textbf{57.83} & \textbf{54.75} \\
    \bottomrule
    \end{tabular}
    }
    \caption{Evaluation of our SFT models (upper part) and existing strong baselines (lower part) on URO-Bench (EN). Und.: Understanding. Conv.: Oral Conversation.}
    \label{tab:uro_bench}
\end{table*}

URO-Bench results (\Cref{tab:uro_bench}) confirm that UltraVoice SFT enhances general conversational skills. All models showed substantial gains across \textit{Understanding}, \textit{Reasoning}, and \textit{Oral Conversation}, with average gains of \textbf{+10.84\%} on Basic and \textbf{+7.87\%} on Pro. Notably, the VocalNet-7B SFT model establishes a new state-of-the-art, outperforming strong baselines like Qwen2.5-Omni-7B~\citep{Qwen2.5-Omni} and GLM4-Voice-9B~\citep{GLM-4-Voice}, highlighting practical value beyond style control. The only exception is a performance drop in Pro Reasoning (from 24.72 to 20.07) for the smallest model, SLAM-Omni-0.5B. We attribute this to the current dataset's focus on single-turn interactions, which may not sufficiently support complex, multi-turn reasoning. Incorporating multi-turn dialogue examples in SFT could address this.

\subsection{Validating Data Quality via Controllable Text-To-Speech}
To validate the quality and utility of our synthesised data, we repurposed it into a controllable TTS dataset. This new dataset, derived from five stylistic dimensions in UltraVoice (speed, volume, emotion, accent, and composite styles), consists of explicit instruction-speech pairs. Following the pipeline of EmoVoice~\citep{EmoVoice}, we fine-tuned the same pre-trained TTS model (EmoVoice-Pre.) on our dataset to ensure a fair comparison.

% \begin{table*}[h]
% \centering
% \small
% \resizebox{\textwidth}{!}{
% \begin{tabular}{l l c c c c}
% \toprule
% \textbf{\makecell{Testset}} & \textbf{Model} & \textbf{WER ↓} & \textbf{Emo\_Sim ↑} & \textbf{Emo\_Recall ↑} & \textbf{UTMOS ↑} \\
% \midrule
% \multirow{5}{*}{\makecell{EmoVoice-DB}} 
%   & PromptTTS       & \textbf{2.11} & 0.87 & 0.29 & 4.32 \\
%   & CosyVoice       & 3.61 & \underline{0.89} & 0.33 & 4.33 \\
%   & CosyVoice2      & 3.61 & 0.86 & \underline{0.37} & \textbf{4.42} \\
%   & EmoVoice-0.5B   & \underline{2.73} & \textbf{0.91} & \textbf{0.40} & \underline{4.36} \\
%   & UltraVoice-0.5B-SFT  & 5.41 & \underline{0.89} & 0.35 & \underline{4.36} \\
% \midrule
% \multirow{3}{*}{\makecell{UltraVoice}}
%   & EmoVoice-0.5B        & 19.82 & \underline{0.94} & \textbf{0.40} & 4.29 \\
%   & EmoVoice-0.5B-Pre--trained      & \underline{14.26} & 0.91 & 0.32 & \textbf{4.49} \\
%   & UltraVoice-0.5B-SFT  & \textbf{3.97}  & \textbf{0.95} & \underline{0.39} & \underline{4.46} \\
% \bottomrule
% \end{tabular}
% }
% \caption{Performance of our \textbf{UltraVoice-0.5B-SFT} model on emotional TTS tasks. The evaluation is conducted on both an out-of-domain test set (EmoVoice-DB, top) and an in-domain test set (UltraVoice, bottom). \textbf{Bold} and \underline{underlined} values denote the best and second-best results, respectively.}
% \label{tab:instruction_tts_results}
% \end{table*}

\begin{table}[htbp] % 1. 去掉星号，改为单栏
\centering
\small
\addtolength{\tabcolsep}{-4.8pt} % 2. 压缩列间距
\resizebox{\columnwidth}{!}{ % 3. 强制缩放到单栏宽度
\begin{tabular}{l l c c c c}
\toprule
\textbf{Test set} & \textbf{Model} & \textbf{WER ↓} & \textbf{E-Sim ↑} & \textbf{E-Rec ↑} & \textbf{UTMOS ↑} \\ % 4. 简写表头
\midrule
\multirow{5}{*}{\makecell{EmoVoice-\\DB}} % 使用 makecell 强制折行
  & PromptTTS       & \textbf{2.11} & 0.87 & 0.29 & 4.32 \\
  & CosyVoice       & 3.61 & \underline{0.89} & 0.33 & 4.33 \\
  & CosyVoice2      & 3.61 & 0.86 & \underline{0.37} & \textbf{4.42} \\
  & EmoVoice   & \underline{2.73} & \textbf{0.91} & \textbf{0.40} & \underline{4.36} \\
  & UltraVoiceTTS   & 5.41 & \underline{0.89} & 0.35 & \underline{4.36} \\ % 5. 简写模型名
\midrule
\multirow{3}{*}{\makecell{UltraVoice}}
  & EmoVoice   & 19.82 & \underline{0.94} & \textbf{0.40} & 4.29 \\
  & EmoVoice-Pre.  & \underline{14.26} & 0.91 & 0.32 & \textbf{4.49} \\
  & UltraVoiceTTS   & \textbf{3.97}  & \textbf{0.95} & \underline{0.39} & \underline{4.46} \\
\bottomrule
\end{tabular}
}
\caption{Performance of our \textbf{UltraVoiceTTS} model on emotional TTS tasks. \textbf{Bold} and \underline{underlined} values denote the best and second-best results, respectively.}
\label{tab:instruction_tts_results}
\end{table}

\textbf{UltraVoiceTTS} demonstrates robust multi-dim\-en\-sional style control. As shown in \Cref{tab:instruction_tts_results}, on emotional control tasks, our model achieves competitive performance against strong baselines such as PromptTTS~\citep{prompttts}, CosyVoice~\citep{CosyVoice1,CosyVoice2}, and EmoVoice~\citep{EmoVoice} on the out-of-domain EmoVoice-DB test set. Crucially, on our in-domain UltraVoice data, it substantially reduces the Word Error Rate (WER) to \textbf{3.97} from 19.82 achieved by EmoVoice, while maintaining high emotional similarity and naturalness. Furthermore, as detailed in \Cref{tab:ultravoice_mos_ifr}, the model consistently improves both MOS and IFR scores across all other tested dimensions (accent, speed, volume, and composite styles) compared to the pre-trained baseline. We omit the fully fine-tuned EmoVoice-0.5B from this broader comparison due to its poor robustness, already indicated by its high WER on our dataset. These results confirm that our instruction-style data effectively enhances controllable synthesis across a diverse range of styles.

\begin{table}[htbp] % 去掉星号
\centering
\small
\addtolength{\tabcolsep}{-5.2pt} % 极度收紧列间距
\resizebox{\columnwidth}{!}{
\begin{tabular}{l ccc ccc ccc ccc ccc}
\toprule
\multirow{2}{*}{\textbf{Model}} & 
\multicolumn{2}{c}{\textbf{Emotion}} & 
\multicolumn{2}{c}{\textbf{Accent}} & 
\multicolumn{2}{c}{\textbf{Speed}} & 
\multicolumn{2}{c}{\textbf{Volume}} & 
\multicolumn{2}{c}{\textbf{Comp.}} \\ % 缩写 Composite
\cmidrule(lr){2-3} \cmidrule(lr){4-5} \cmidrule(lr){6-7} \cmidrule(lr){8-9} \cmidrule(lr){10-11}
& MOS & IFR & MOS & IFR & MOS & IFR & MOS & IFR & MOS & IFR \\
\midrule
EmoVoice-Pre. & 2.52 & 50.29 & 3.62 & 74.67 & 4.60 & 95.33 & 3.92 & 77.33 & 3.59 & 76.00 \\
UltraVoiceTTS & 3.08 & 67.43 & 4.10 & 88.33 & 4.74 & 98.67 & 4.28 & 85.33 & 3.92 & 86.00 \\
\bottomrule
\end{tabular}
}
\caption{MOS and IFR results of UltraVoiceTTS across five style dimensions.}
\label{tab:ultravoice_mos_ifr}
\end{table}

\section{Conclusion}

% In this work, we introduce \textbf{UltraVoice}, the first large-scale speech dialogue dataset engineered for multiple fine-grained speech style control. By fine-tuning mainstream spoken dialogue models on UltraVoice, we significantly enhanced their controllability over diverse fine-grained speech styles, while also improving their overall speech naturalness and general conversational competence. The dataset's high quality and generalization were further validated through strong performance on the URO-Bench benchmark and in controllable TTS tasks, establishing a solid foundation for expressive spoken dialogue modeling. While this work represents a significant step forward, the full scope of human-like expressive speech presents formidable long-term challenges. The framework and data we provide can be extended to explore these frontiers, such as modeling the dynamic evolution of styles within multi-turn conversations or capturing the nuanced paralinguistic features in massively multilingual contexts. Addressing these complex scenarios, though beyond the scope of this paper, will be critical for developing the next generation of truly context-aware and intelligent speech interaction systems.

We introduce \textbf{UltraVoice}, the first large-scale spoken dialogue dataset for multiple fine-grained speech style control. By fine-tuning mainstream spoken dialogue models on UltraVoice, we significantly enhanced their controllability over diverse fine-grained speech styles, while also improving their overall speech naturalness and general conversational competence. The dataset's high quality and generalization were validated through strong performance on the URO-Bench benchmark and in controllable TTS tasks, establishing a solid foundation for expressive spoken dialogue modeling.
\section*{Limitations}

Although Audio-Language Models enable scalable evaluation and align well with human judgments, expressive speech requires more specialized metrics. Moreover, UltraVoice has limited multilingual coverage and primarily targets single-turn dialogues. Broader multilingual instructions and complex multi-turn interactions remain important directions for future work.


\section*{Potential Risks}

% [original]
% The primary ethical risk of UltraVoice involves its potential for malicious exploitation. By providing over 830 hours of speech data with fine-grained control over six stylistic dimensions including emotion, speed, volume, accent, language, and composite styles, this work could lower the technical barrier for generating highly realistic and persuasive synthetic audio. UltraVoice is intended strictly for advancing research in expressive and natural spoken dialogue models. We strongly advocate for the concurrent development of robust audio watermarking and deepfake detection technologies to mitigate these potential harms.

% [revised]
UltraVoice could be misused to generate realistic speech for impersonation, fraud, or deception. It is intended solely for research. We encourage responsible use alongside audio watermarking, provenance tracking, and deepfake detection to mitigate these risks.

\section*{Acknowledgments}
This work was supported by the National Natural Science Foundation of China (No. 62376031, No. 62671373 and No. U23B2018), the Science and Technology Innovation (STI) 2030-Major Project (2022ZD0208700), and Yangtze River Delta Science and Technology Innovation Community Joint Research Project (2024CSJGG1100).

\bibliography{custom}

\newpage
% \section*{Appendix}
\appendix

\section{Human Subjective Evaluation}
\label{app:human_eval}

% To verify the reliability of our ALM-based evaluation framework, we conducted a targeted human study. Given the large scale of our full test set, which contains 2,300 samples, we selected SLAM-Omni-0.5B as a representative case. We randomly sampled 115 test samples, with 5 samples from each of the 23 fine-grained sub-dimensions. We recruited 15 human experts to independently provide Mean Opinion Score (MOS) and Instruction Following Rate (IFR) ratings following the same evaluation criteria as Gemini-2.5-Flash.

% As shown in Tables~\ref{tab:human_eval_mos} and~\ref{tab:human_eval_ifr}, the performance gains observed by human experts are highly consistent with those reported by the ALM evaluator. At the sample level, Gemini-2.5-Flash also shows positive Spearman correlations with human judgments, reaching $\rho=0.6139$ for MOS and $\rho=0.6482$ for IFR.

To verify our ALM-based evaluation framework, we conducted a human subjective evaluation on SLAM-Omni-0.5B as a representative case. We randomly sampled 115 test samples, with 5 from each of the 23 fine-grained sub-dimensions, and asked 15 experts to rate MOS and IFR using the same criteria as Gemini-2.5-Flash. The experts were recruited from relevant academic and industry communities and compensated at standard market rates for their annotation work. Inter-annotator agreement was generally moderate to substantial, with an overall Fleiss' $\kappa$ of 0.730 for IFR and an overall Krippendorff's $\alpha$ of 0.761 for MOS.

As shown in~\Cref{tab:human_eval_mos} and~\Cref{tab:human_eval_ifr}, the performance gains observed by human experts are highly consistent with those reported by the ALM evaluator. At the sample level, Gemini-2.5-Flash also shows positive Spearman correlations with human judgments, reaching $\rho=0.61$ for MOS and $\rho=0.65$ for IFR. These results further support the use of ALM-based evaluation in our experiments.

% To verify our ALM-based evaluation framework, we conducted a human subjective evaluation on SLAM-Omni-0.5B as a representative case. Since the full UltraVoice test set contains 2,300 samples and exhaustive human annotation is costly, we constructed a balanced subset by sampling 115 test samples, with 5 samples from each of the 23 fine-grained sub-dimensions, and asked 15 experts to rate MOS and IFR using the same criteria as Gemini-2.5-Flash.

% As shown in~\Cref{tab:human_eval_mos} and~\Cref{tab:human_eval_ifr}, human judgments show consistent gains after SFT in both MOS and IFR, aligning with the ALM-based results. Gemini-2.5-Flash also exhibits positive sample-level Spearman correlations with human judgments, reaching $\rho=0.6139$ for MOS and $\rho=0.6482$ for IFR. These results further support the use of ALM-based evaluation in our experiments.

\begin{table}[h]
\centering
\small
\setlength{\tabcolsep}{3.5pt}
\begin{tabular}{lccc ccc}
\toprule
\multirow{2}{*}{Dim.} 
& \multicolumn{3}{c}{Gemini-2.5-Flash} 
& \multicolumn{3}{c}{Human} \\
\cmidrule(lr){2-4}\cmidrule(lr){5-7}
& Base & SFT & $\Delta$ (\%) & Base & SFT & $\Delta$ (\%) \\
\midrule
Speed & 3.25 & 3.88 & +19.38 & 3.31 & 4.47 & +35.05 \\
Lang. & 1.15 & 1.15 & 0.00   & 1.34 & 2.44 & +82.09 \\
Vol.  & 2.51 & 3.34 & +33.07 & 2.95 & 3.90 & +32.20 \\
Emo.  & 2.42 & 3.30 & +36.36 & 3.04 & 3.99 & +31.25 \\
Acc.  & 1.94 & 3.80 & +95.88 & 2.63 & 4.30 & +63.50 \\
Comp. & 2.22 & 3.32 & +49.55 & 2.99 & 3.51 & +17.39 \\
\midrule
Avg.  & 2.24 & 3.23 & +44.20 & 2.73 & 3.90 & +42.86 \\
\bottomrule
\end{tabular}
\caption{
MOS comparison between Gemini-2.5-Flash and human evaluation. $\Delta$ (\%) denotes relative improvement.
}
\label{tab:human_eval_mos}
\end{table}

\begin{table}[h]
\centering
\small
\setlength{\tabcolsep}{3.5pt}
\begin{tabular}{lccc ccc}
\toprule
\multirow{2}{*}{Dim.} 
& \multicolumn{3}{c}{Gemini-2.5-Flash} 
& \multicolumn{3}{c}{Human} \\
\cmidrule(lr){2-4}\cmidrule(lr){5-7}
& Base & SFT & $\Delta$ & Base & SFT & $\Delta$ \\
\midrule
Speed & 60.00 & 100.00 & +40.00 & 42.67 & 95.56 & +52.89 \\
Lang. & 0.00  & 40.00  & +40.00 & 0.00  & 48.89 & +48.89 \\
Vol.  & 33.33 & 73.33  & +40.00 & 37.78 & 86.22 & +48.44 \\
Emo.  & 17.14 & 62.86  & +45.72 & 22.67 & 82.48 & +59.81 \\
Acc.  & 30.00 & 96.67  & +66.67 & 27.11 & 92.67 & +65.56 \\
Comp. & 0.00  & 40.00  & +40.00 & 14.67 & 40.00 & +25.33 \\
\midrule
Avg.  & 25.22 & 73.91  & +48.69 & 25.10 & 81.10 & +56.00 \\
\bottomrule
\end{tabular}
\caption{
IFR comparison between Gemini-2.5-Flash and human evaluation. $\Delta$ denotes absolute improvement in percentage points.
}
\label{tab:human_eval_ifr}
\end{table}

% \begin{table}[h]
% \centering
% \small
% \setlength{\tabcolsep}{10pt}
% \begin{tabular}{lc}
% \toprule
% Metric & Spearman's $\rho$ \\
% \midrule
% MOS & 0.6139 \\
% IFR & 0.6482 \\
% \bottomrule
% \end{tabular}
% \caption{
% Sample-level Spearman correlation between Gemini-2.5-Flash and human judgments.
% }
% \label{tab:human_eval_corr}
% \end{table}

\section{The Detailed Performance Comparison}\label{app:detailed_performance_comparison}
The detailed performance corresponding to~\Cref{fig:instr_acc} is presented in~\Cref{tab:instruction_following_rate}.

\begin{table}[H] % 1. 去掉星号，使用 [H] 固定位置（需 \usepackage{float}）
\centering
\small
\addtolength{\tabcolsep}{-4.5pt} % 2. 进一步收紧列间距以适应单栏
\resizebox{\columnwidth}{!}{ % 3. 强制缩放到单栏宽度
\begin{tabular}{lccccccc}
\toprule
\textbf{Model} & \textbf{Speed} & \textbf{Lang.} & \textbf{Vol.} & \textbf{Emo.} & \textbf{Acc.} & \textbf{Desc.} & \textbf{Over.} \\ % 4. 建议缩写表头
\midrule
\textbf{GPT-4o (GT)} & 96.50 & 92.33 & 89.67 & 76.32 & 96.33 & 76.00 & 87.68 \\
\cmidrule(lr){1-8}
SLAM-Omni-0.5B Base & 50.67 & 0.00 & 41.00 & 13.86 & 42.67 & 23.00 & 28.30 \\
SLAM-Omni-0.5B SFT  & 92.00 & 30.33 & 70.33 & 58.71 & 88.33 & 54.00 & 68.39 \\
$\Delta$     & \textcolor{teal}{+41.33} & \textcolor{teal}{+30.33} & \textcolor{teal}{+29.33} 
                   & \textcolor{teal}{+44.86} & \textcolor{teal}{+45.67} & \textcolor{teal}{+31.00} 
                   & \textcolor{teal}{+40.09} \\
\cmidrule(lr){1-8}
VocalNet-1B Base & 71.67 & 0.33 & 54.67 & 19.00 & 49.08 & 27.00 & 36.28 \\
VocalNet-1B SFT  & 89.00 & 0.33 & 65.00 & 50.00 & 68.67 & 61.00 & 55.91 \\
$\Delta$               & \textcolor{teal}{+17.33} & \textcolor{gray}{+0.00} & \textcolor{teal}{+10.33} 
                          & \textcolor{teal}{+31.00} & \textcolor{teal}{+19.58} & \textcolor{teal}{+34.00} 
                          & \textcolor{teal}{+19.64} \\
\cmidrule(lr){1-8}
VocalNet-7B Base & 75.33 & 24.00 & 50.67 & 22.43 & 52.67 & 27.00 & 41.30 \\
VocalNet-7B SFT  & 90.67 & 34.00 & 63.00 & 46.43 & 59.00 & 45.00 & 55.96 \\
$\Delta$            & \textcolor{teal}{+15.33} & \textcolor{teal}{+10.00} & \textcolor{teal}{+12.33} 
                         & \textcolor{teal}{+24.00} & \textcolor{teal}{+6.33} & \textcolor{teal}{+18.00} 
                         & \textcolor{teal}{+14.65} \\
\cmidrule(lr){1-8}
VocalNet-8B Base & 72.33 & 0.67 & 54.67 & 28.43 & 69.83 & 28.00 & 44.74 \\
VocalNet-8B SFT  & 90.33 & 1.00 & 69.00 & 56.43 & 70.67 & 65.00 & 59.35 \\
$\Delta$             & \textcolor{teal}{+18.00} & \textcolor{teal}{+0.33} & \textcolor{teal}{+14.33} 
                          & \textcolor{teal}{+28.00} & \textcolor{teal}{+0.83} & \textcolor{teal}{+37.00} 
                          & \textcolor{teal}{+14.61} \\
\bottomrule
\end{tabular}
}
\caption{Detailed IFR (\%) results across six fine-grained speech style control dimensions for each model.}
\label{tab:instruction_following_rate}
\end{table}

\section{Supervised Fine-tuning Details}\label{app:sft_config} 

This section details the configurations used for the Supervised Fine-tuning (SFT) of the Spoken Dialogue and Controllable TTS models, as mentioned in our experiments (\Cref{sec:experiment}).

% --- 表格 1：SLAM-Omni ---
\begin{table}[H] % 使用大写 H
\centering
\small
\begin{tabular}{lc}
\toprule
\textbf{Parameter} & \textbf{Value} \\
\midrule
Batch Size & 1 \\
Gradient Accumulation Steps & 1 \\
Learning Rate & $1 \times 10^{-5}$ \\
Training Epochs & 5 \\
Context Length & 4,096 \\
Hardware & 4 $\times$ NVIDIA A100-80G \\
Learning Rate Scheduler & Linear \\
Optimiser & AdamW \\
Warmup Steps & 5,000 \\
Weight Decay & 0.0 \\
Use FP16 & True \\
\bottomrule
\end{tabular}
\caption{SFT Training Configuration for SLAM-Omni-0.5B-SFT.}
\label{tab:slamomni_sft_config}
\end{table}

% --- 表格 2：VocalNet ---
\begin{table}[H]
\centering
\small
\begin{tabular}{lc}
\toprule
\textbf{Parameter} & \textbf{Value} \\
\midrule
Batch Size & 4 \\
Gradient Accumulation Steps & 4 \\
Learning Rate & $5 \times 10^{-5}$ \\
Training Epochs & 3 \\
Context Length & 4,096 \\
Hardware & 4 $\times$ NVIDIA A100-80G \\
Learning Rate Scheduler & Cosine \\
Optimiser & AdamW \\
Warmup Ratio & 0.03 \\
Weight Decay & 0.0 \\
Use BF16 & True \\
\bottomrule
\end{tabular}
\caption{SFT Training Configuration for VocalNet1B /7B/8B-SFT.}
\label{tab:vocalnet_sft_config}
\end{table}

% --- 表格 3：UltraVoice ---
\begin{table}[H]
\centering
\small
\begin{tabular}{lc}
\toprule
\textbf{Parameter} & \textbf{Value} \\
\midrule
Batch Size & 6 \\
Gradient Accumulation Steps & 1 \\
Learning Rate & $1 \times 10^{-5}$ \\
Training Epochs & 400 \\
Context Length & 4,096 \\
Hardware & 4 $\times$ NVIDIA A100-80G \\
Learning Rate Scheduler & Linear \\
Optimiser & AdamW \\
Warmup Steps & 1,000 \\
Weight Decay & 0.0 \\
Use FP16 & True \\
\bottomrule
\end{tabular}
\caption{SFT Training Configuration for UltraVoice-0.5B-SFT.}
\label{tab:ultravoice_sft_config}
\end{table}

% --- 表格 4：模型总览 (跨栏显示建议) ---
% 如果这个表太宽，建议使用 table* 环境使其横跨两栏
\begin{table}[H]
\centering
\addtolength{\tabcolsep}{-2pt} 
\resizebox{\columnwidth}{!}{
\begin{tabular}{lcccc}
\toprule
\textbf{Model Name} & \textbf{Speech Encoder} & \textbf{LLM Backbone} & \textbf{Size} & \textbf{Decoder} \\
\midrule
SLAM-Omni-0.5B & Whisper-small-v3 & Qwen2 & 0.5B & CosyVoice1 \\
VocalNet-1B & Whisper-large-v3 & LLaMA3.2 & 1B & CosyVoice2 \\
VocalNet-7B & Whisper-large-v3 & Qwen2.5 & 7B & CosyVoice2 \\
VocalNet-8B & Whisper-large-v3 & LLaMA3.1 & 8B & CosyVoice2 \\
\bottomrule
\end{tabular}
}
\caption{Spoken dialogue model configurations for SFT experiments.}
\label{tab:model-settings}
\end{table}

\section{Detailed Dataset Statistics}\label{app:detailed_datasest_statistics}
This section provides detailed statistics for each fine-grained control sub-dimension in UltraVoice.

\begin{table}[h] % 去掉了 *，使其变为单栏
\centering
\small % 降低基础字号
\addtolength{\tabcolsep}{-3.5pt} % 紧凑列间距，视情况调整数值
\resizebox{\columnwidth}{!}{ % 强制适配栏宽
\begin{tabular}{llcccc}
\toprule
\textbf{Dimension} & \textbf{\makecell{Sub\\Dimension}} & \textbf{\#Cnt.} & \textbf{Dur.(h)} & \textbf{CER} & \textbf{UTMOS} \\
\midrule
\multirow{7}{*}{Emotion} & Angry & 3097 & 26.23 & 7.15 & 4.01 \\
                         & Disgusted & 3032 & 26.03 & 5.83 & 3.97 \\
                         & Fearful & 2590 & 23.49 & 6.74 & 3.83 \\
                         & Happy & 3097 & 27.05 & 6.14 & 4.05 \\
                         & Neutral & 3848 & 31.75 & 4.55 & 4.05 \\
                         & Sad & 2147 & 19.40 & 5.01 & 3.86 \\
                         & Surprised & 3398 & 28.59 & 7.75 & 4.03 \\
\midrule
\multirow{3}{*}{Volume} & High & 3575 & 29.54 & 5.98 & 4.05 \\
                        & Low & 3622 & 30.10 & 5.07 & 3.27 \\
                        & Normal & 3957 & 31.73 & 4.87 & 4.07 \\
\midrule
\multirow{3}{*}{Speed} & Fast & 4370 & 34.48 & 5.22 & 4.04 \\
                       & Normal & 3864 & 31.32 & 4.59 & 4.06 \\
                       & Slow & 2100 & 19.48 & 4.51 & 4.05 \\
\midrule
\multirow{6}{*}{Accent} & AU & 4683 & 43.89 & 7.27 & 4.09 \\
                        & CA & 4844 & 45.41 & 6.63 & 4.08 \\
                        & GB & 4953 & 46.10 & 5.19 & 4.12 \\
                        & IN & 4128 & 39.62 & 5.75 & 4.05 \\
                        & SG & 3702 & 35.37 & 9.38 & 4.06 \\
                        & ZA & 4529 & 42.92 & 6.45 & 4.05 \\
\midrule
\multirow{3}{*}{Language} & Chinese & 4388 & 35.83 & 5.60 & 3.85 \\
                          & Japanese & 2468 & 22.19 & 10.81 & 3.99 \\
                          & Korean & 4297 & 35.81 & 5.79 & 3.99 \\
\midrule
Composite  & EN & 4143 & 33.47 & 5.02 & 3.97 \\
\midrule
General QA & EN & 15938 & 93.12 & 6.69 & 4.15 \\
\bottomrule
\end{tabular}
}
\caption{Detailed statistics for each fine-grained control dimension, including sample count (\textbf{\#Cnt.}), duration in hours (\textbf{Dur.(h)}), \textbf{CER}, and \textbf{UTMOS}.}
\label{tab:detail_data_statistics}
\end{table}

\begin{table}[h] % 移除 * 变为单栏
\centering
\footnotesize % 使用较小字号
\setlength{\tabcolsep}{2.2pt} % 极致压缩列间距
\resizebox{\columnwidth}{!}{ % 确保不溢出
\begin{tabular}{ll ccc ccc}
\toprule
\multirow{2}{*}{\textbf{Dimension}} & \multirow{2}{*}{\textbf{\makecell{Sub\\Dim.}}} & \multicolumn{3}{c}{\textbf{Dur. (s)}} & \multicolumn{3}{c}{\textbf{Words}} \\
\cmidrule(lr){3-5} \cmidrule(lr){6-8}
& & \textbf{Dia.} & \textbf{Ins.} & \textbf{Res.} & \textbf{Dia.} & \textbf{Ins.} & \textbf{Res.} \\
\midrule
\multirow{7}{*}{Emotion} & Angry & 30.49 & 11.20 & 19.30 & 71.43 & 30.23 & 41.20 \\
                         & Disgusted & 30.90 & 10.72 & 20.18 & 65.08 & 29.07 & 36.01 \\
                         & Fearful & 32.65 & 10.95 & 21.70 & 72.81 & 28.98 & 43.83 \\
                         & Happy & 31.44 & 11.04 & 20.40 & 72.81 & 30.20 & 42.62 \\
                         & Neutral & 29.70 & 11.32 & 18.38 & 66.55 & 29.45 & 37.10 \\
                         & Sad & 32.52 & 10.08 & 22.44 & 64.38 & 26.94 & 37.44 \\
                         & Surprised & 30.29 & 11.17 & 19.12 & 68.29 & 29.27 & 39.02 \\
\midrule
\multirow{3}{*}{Volume} & High & 29.74 & 11.44 & 18.30 & 67.19 & 30.61 & 36.58 \\
                        & Low & 29.91 & 11.47 & 18.44 & 64.95 & 30.90 & 34.05 \\
                        & Normal & 28.87 & 11.56 & 17.31 & 65.95 & 30.56 & 35.39 \\
\midrule
\multirow{3}{*}{Speed}  & Fast & 28.40 & 12.61 & 15.79 & 75.48 & 34.36 & 41.12 \\
                        & Normal & 29.18 & 11.39 & 17.80 & 67.66 & 30.37 & 37.28 \\
                        & Slow & 33.39 & 10.67 & 22.72 & 62.83 & 28.42 & 34.41 \\
\midrule
\multirow{6}{*}{Accent} & AU & 33.74 & 13.88 & 19.86 & 83.10 & 36.75 & 46.35 \\
                        & CA & 33.75 & 13.84 & 19.91 & 87.24 & 37.41 & 49.83 \\
                        & GB & 33.51 & 14.39 & 19.11 & 84.47 & 38.11 & 46.37 \\
                        & IN & 34.55 & 13.03 & 21.52 & 80.00 & 34.89 & 45.11 \\
                        & SG & 34.39 & 13.17 & 21.22 & 80.20 & 35.28 & 44.91 \\
                        & ZA & 34.12 & 13.57 & 20.55 & 83.72 & 36.67 & 47.05 \\
\midrule
\multirow{3}{*}{Language}& Chinese & 29.40 & 12.65 & 16.75 & 99.79 & 32.11 & 67.68 \\
                         & Japanese & 32.37 & 12.52 & 19.85 & 70.95 & 31.17 & 39.78 \\
                         & Korean & 30.00 & 12.32 & 17.68 & 114.80 & 32.23 & 82.57 \\
\midrule
Composite & EN & 29.09 & 8.21 & 20.87 & 63.06 & 22.94 & 40.11 \\
\midrule
General QA & EN & 21.03 & 5.06 & 15.97 & 58.03 & 14.47 & 43.56 \\
\bottomrule
\end{tabular}
}
\caption{Detailed statistics: mean duration (\textbf{Dur.}) and word count (\textbf{Words}) for full dialogue (\textbf{Dia.}), instruction (\textbf{Ins.}), and response (\textbf{Res.}).}
\label{tab:detail_duration_word_cnt_statistics}
\end{table}

\newcommand{\subfigwidth}{0.31\textwidth}
\begin{figure*}[htbp!]
    \centering
    
    \begin{subfigure}{\subfigwidth}
        \includegraphics[width=\linewidth]{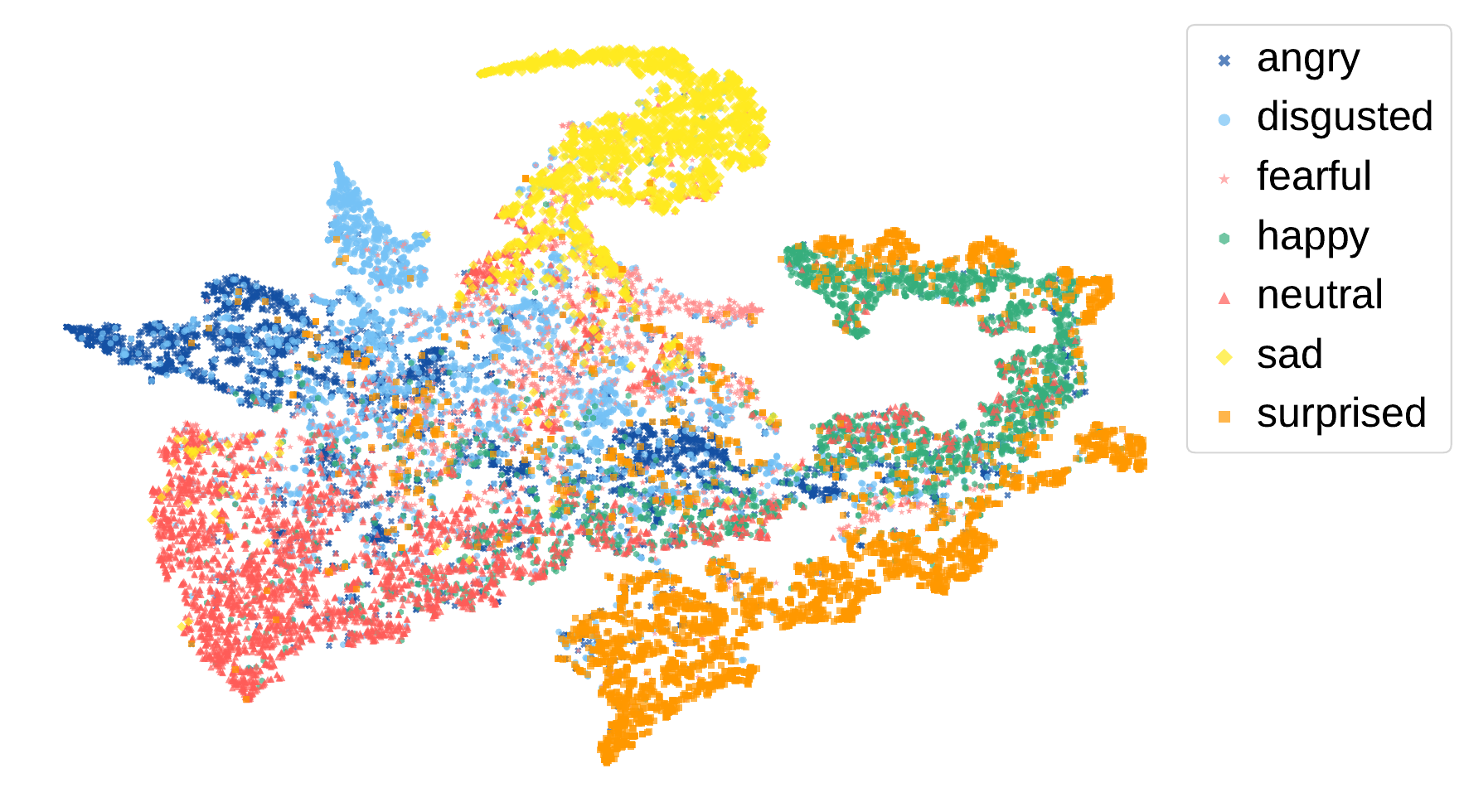}
        \caption{Emotion}
        \label{fig:emotion_tsne}
    \end{subfigure}
    \hfill
    \begin{subfigure}{\subfigwidth}
        \includegraphics[width=\linewidth]{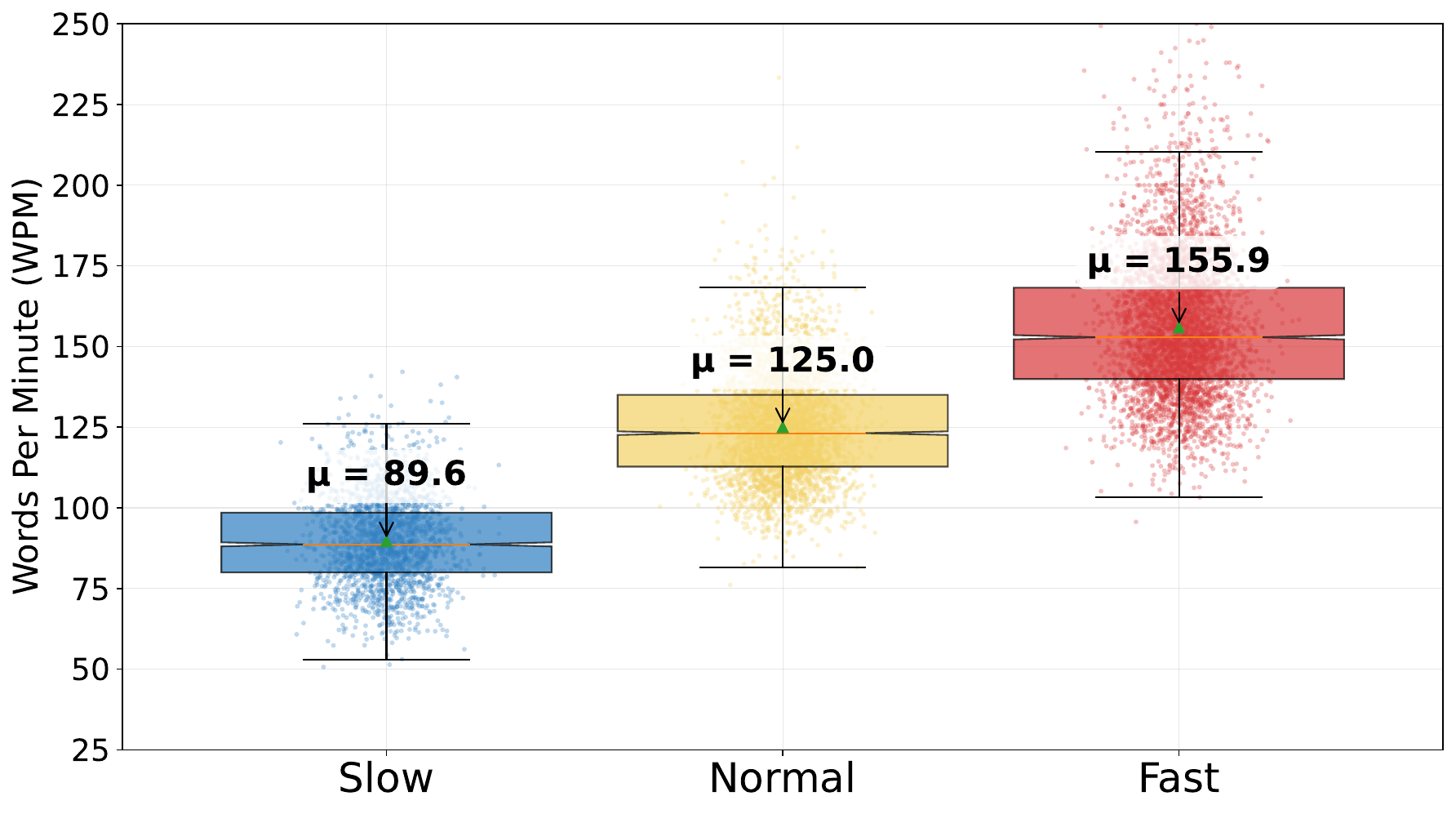}
        \caption{Speed}
        \label{fig:speed_dist}
    \end{subfigure}
    \hfill
    \begin{subfigure}{\subfigwidth}
        \includegraphics[width=\linewidth]{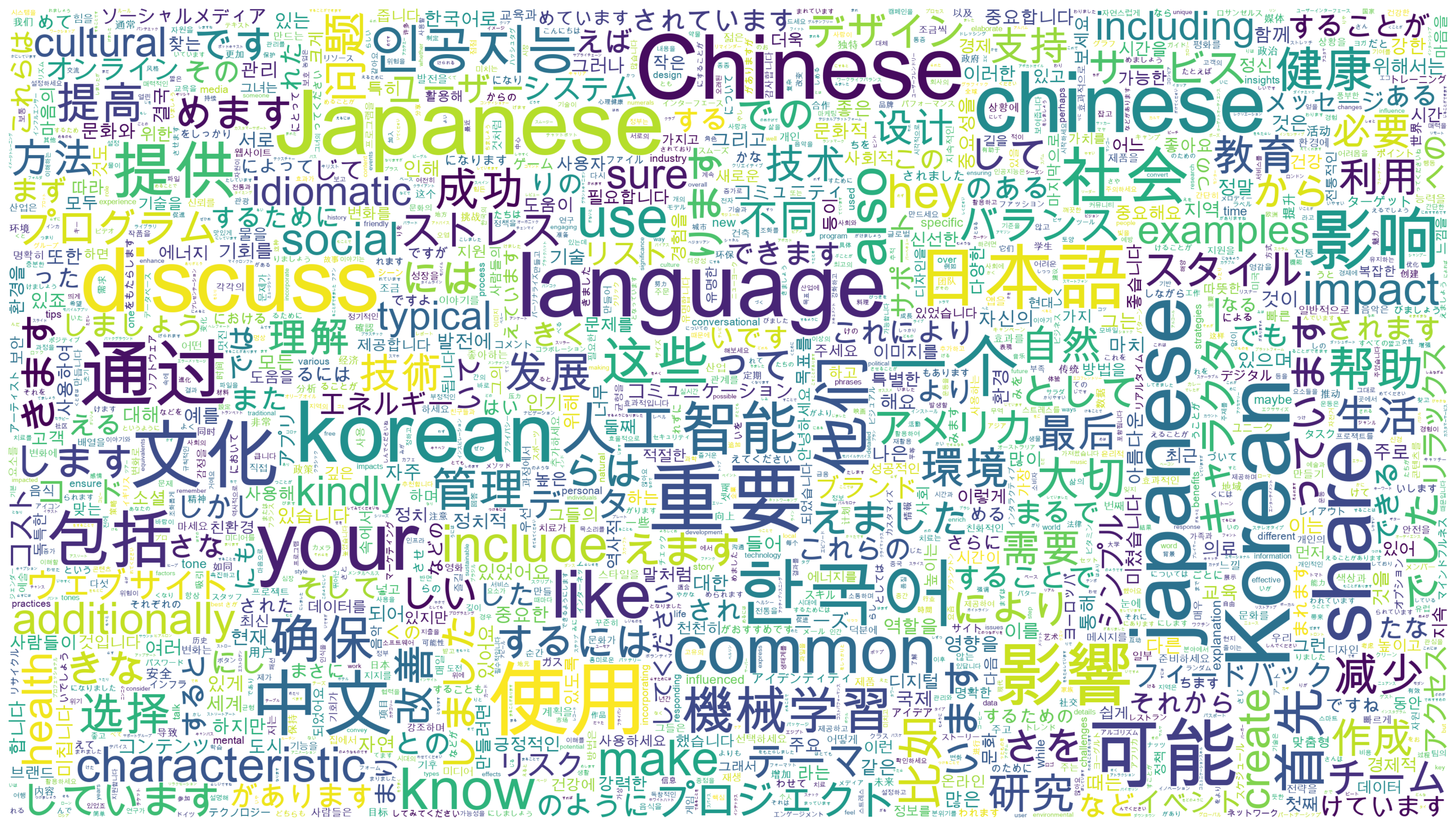}
        \caption{Language}
        \label{fig:lang_word_cloud}
    \end{subfigure}
    
    \vspace{1em}
    
    \begin{subfigure}{\subfigwidth}
        \includegraphics[width=\linewidth]{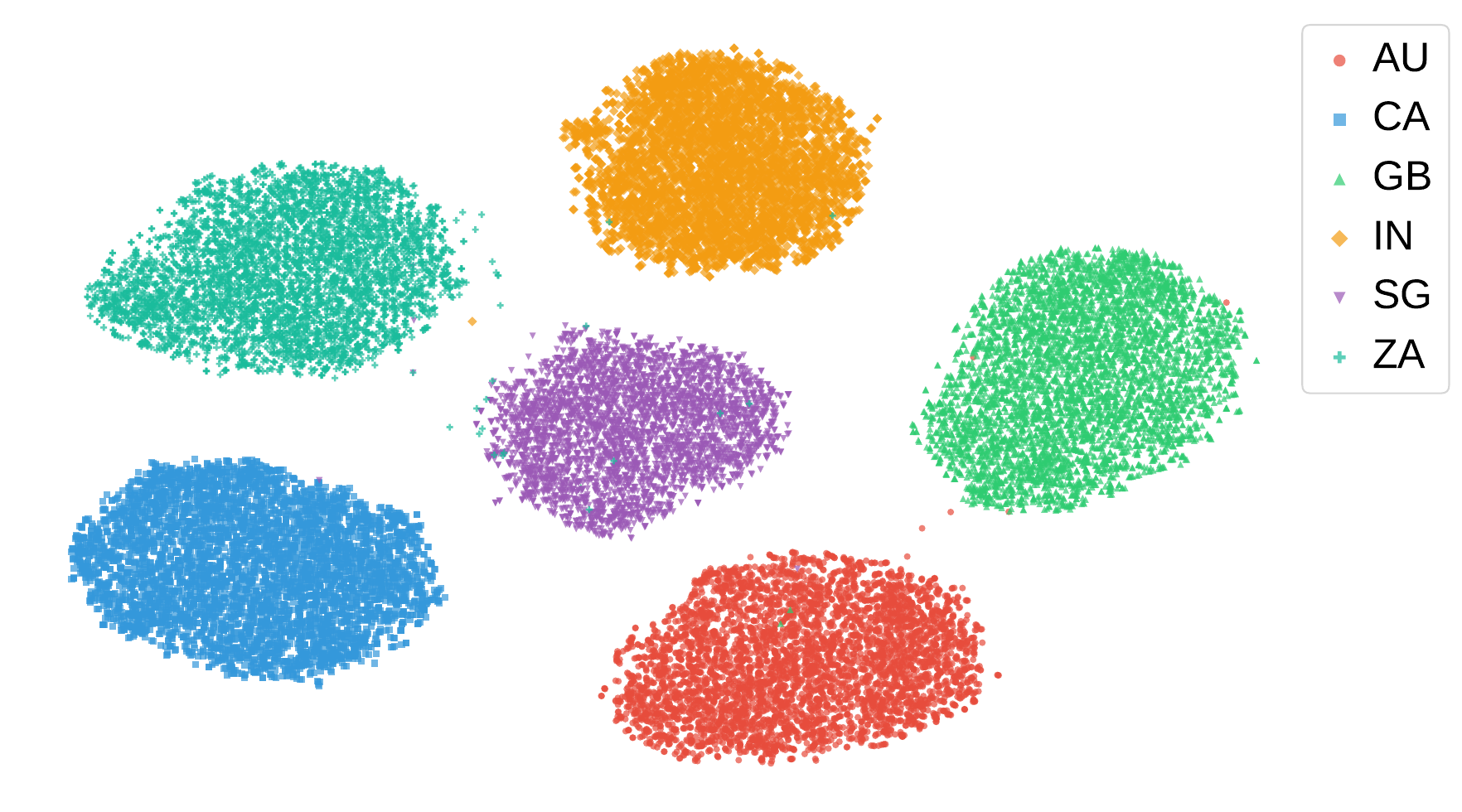}
        \caption{Accent}
        \label{fig:accent_tsne}
    \end{subfigure}
    \hfill
    \begin{subfigure}{\subfigwidth}
        \includegraphics[width=\linewidth]{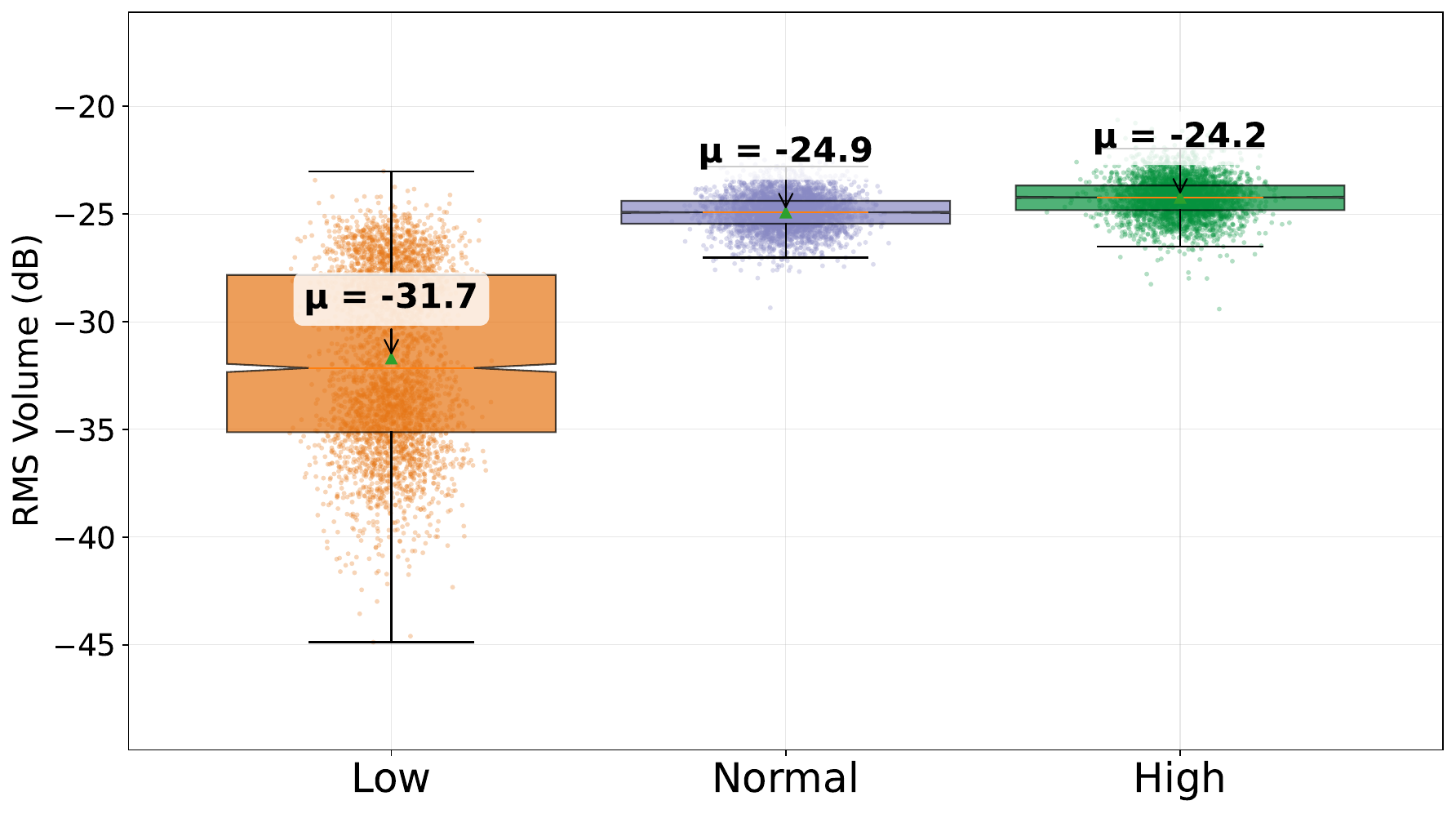}
        \caption{Volume}
        \label{fig:volume_dist}
    \end{subfigure}
    \hfill
    \begin{subfigure}{\subfigwidth}
        \includegraphics[width=\linewidth]{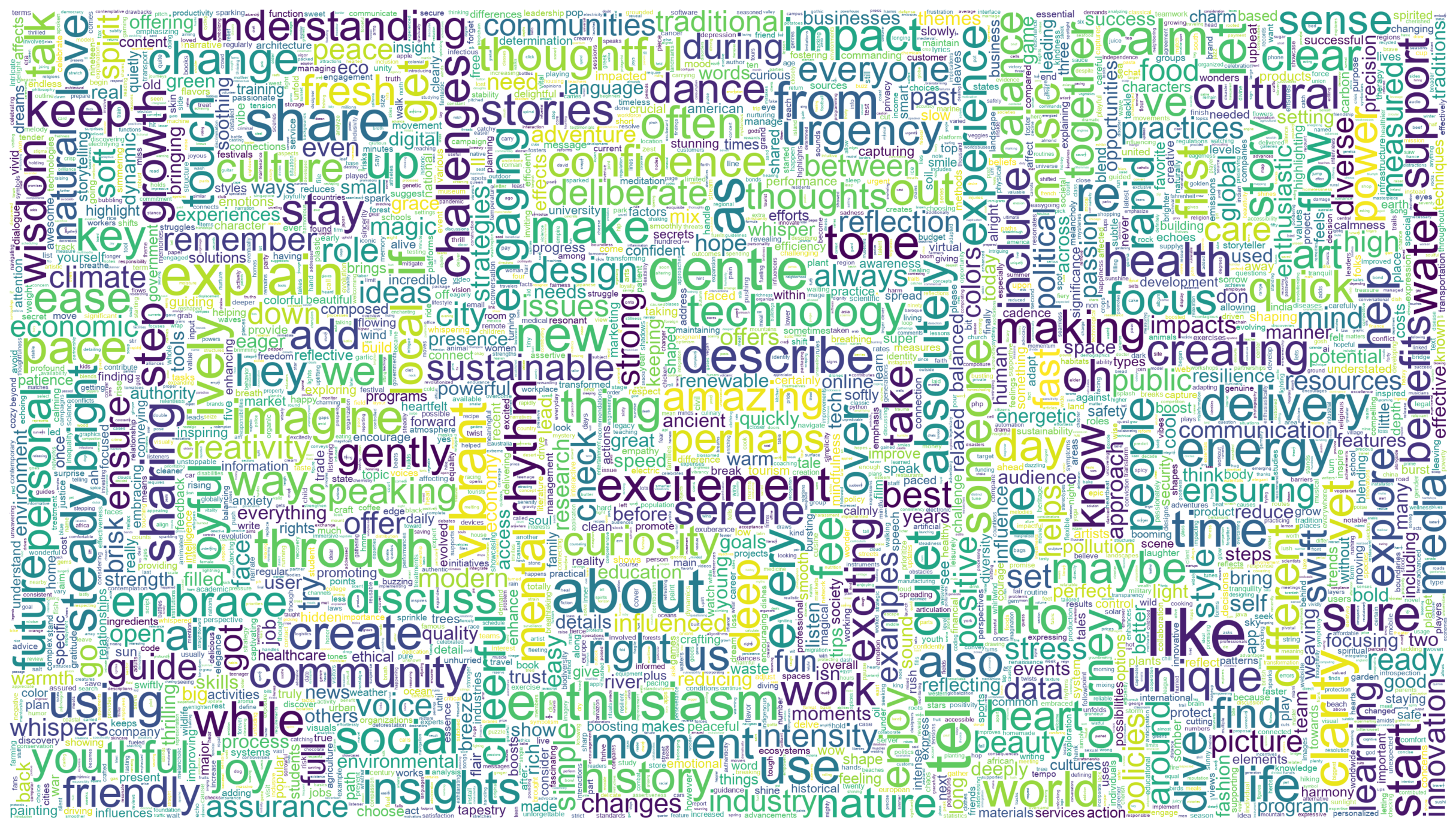}
        \caption{Composite}
        \label{fig:desc_word_cloud}
    \end{subfigure}

   \caption{Statistical visualizations of the six fine-grained speech style control dimensions in UltraVoice. The visualization methods are tailored to the nature of each dimension: t-SNE plots for categorical attributes (Emotion, Accent) demonstrate clear class separability; distributions of physical metrics (Speed, Volume) confirm precise control over acoustic properties; and word clouds (Language, Composite) highlight lexical diversity and expressive richness.}
    \label{fig:quality_assessment}
\end{figure*}

\section{Statement For The Use of Large Language Models (LLMs)}\label{app:llm_usage}

During this work, we utilized LLMs to assist in several aspects of the writing and presentation process. Specifically, LLMs were used for grammar and language refinement, including proofreading for grammatical errors, spelling mistakes, and awkward phrasing; for code correction and debugging, including identifying syntax errors, logical issues, and possible code optimizations; and for assistance in figure creation, including generating plotting scripts and suggesting effective visual representations of data and concepts. These uses were intended to improve the clarity, readability, and accessibility of the manuscript without altering its scientific substance.

The core scientific contributions of this work, including the research ideas, experimental design, data analysis, and final conclusions, are entirely the work of the authors. The role of LLMs was strictly limited to that of an assistive tool to enhance the presentation and accuracy of this work. All outputs generated with the assistance of LLMs were carefully reviewed, edited, and verified by the authors before inclusion in the final manuscript.

\newpage
\section{Prompts}\label{app:prompts}

\subsection{Instruction Rewriting}
% Rewritten Question Generation Prompt
\begin{tcolorbox}[colback=gray!10, colframe=black,width=\linewidth,
                 arc=1mm, auto outer arc, boxrule=0.5pt,
                 breakable]
\textbf{Instruction:} \\
Below is an instruction data for rewriting a user-provided instruction into a speech-oriented question for training a speech-based LLM. Please follow these updated requirements:

\begin{enumerate}[leftmargin=1.5em]
    \item \textbf{Non-Human Language Request Transformation}
        \begin{itemize}[leftmargin=1.5em]
            \item If the user’s request involves technical or non-human language (such as code, formulas, or other specialized terms), rephrase it into a more approachable and human-friendly request. For example:
                \begin{itemize}[leftmargin=1.5em]
                    \item ``Could you explain what this code does?''
                    \item ``How would you describe this formula in simpler terms?''
                \end{itemize}
        \end{itemize}

    \item \textbf{Non-verbal Request Transformation}
        \begin{itemize}[leftmargin=1.5em]
            \item For non-verbal requests, such as those asking to write a piece of text or to perform a task, convert them into action-based expressions using verbs like ``tell'', ``speak'', ``describe'', etc.
        \end{itemize}

    \item \textbf{Incorporating Conversational Fillers}
        \begin{itemize}[leftmargin=1.5em]
            \item Use fillers sparingly to avoid making the question sound unnatural. Ensure the question remains concise.
            \item Add fillers as appropriate (but not too many ``well,'' ``hmm,'' or ``you know'', etc).
        \end{itemize}
   
    \item \textbf{Clarity and Conciseness}
        \begin{itemize}[leftmargin=1.5em]
            \item The question should be relatively brief without excessive verbiage.
            \item Rewrite the instruction into a neutral, clear question suitable for speech input.
        \end{itemize}

    \item \textbf{Number Conversion}
        \begin{itemize}[leftmargin=1.5em]
            \item Convert all numerals into their English word equivalents (e.g., ``six'' instead of ``6,'' ``twenty-two'' instead of ``22'').
            \item This enhances the natural, conversational flow of the question.
        \end{itemize}
\end{enumerate}
   
[instruction]: \{instruction\}

\vspace{1em}

Please output the result in the following JSON format:
\begin{verbatim}
{
    "question_text": {{question_text}}
}
\end{verbatim}
\end{tcolorbox}

\subsection{Response Generation}
% Response Generation Prompt
\begin{tcolorbox}[colback=gray!10, colframe=black,width=\linewidth,
                 arc=1mm, auto outer arc, boxrule=0.5pt,
                 breakable]
\textbf{Instruction:} \\
Below is the transcribed text of a user’s speech query. Please provide a response to this question, which will be converted to speech using TTS. Please follow these requirements for your response:

\begin{enumerate}[leftmargin=1.5em]
    \item Your response should not contain content that cannot be synthesized by the TTS model, such as parentheses, ordered lists, etc. Numbers should be written in English words rather than Arabic numerals.
    \item Your response should be very concise and to the point, avoiding lengthy explanations.
    \item If a specific dialect or style is requested, please incorporate the unique characteristics of that dialect or style into your response text.
    \item Keep your response short enough to generate speech within fifteen to thirty seconds.
\end{enumerate}

[instruction]: \{instruction\}

\vspace{1em}

Please output in JSON format as follows:
\begin{verbatim}
{{
    "response": {{response}}
}}
\end{verbatim}
\end{tcolorbox}

% \newpage
\subsection{Emotion Control}
% Emotion Control Instruction Generation Prompt
\begin{tcolorbox}[colback=gray!10, colframe=black,width=\linewidth,
                 arc=1mm, auto outer arc, boxrule=0.5pt,
                 breakable]
\textbf{Instruction:} \\
Below is an instruction to transform a user-provided conversational question text into an instruction text that integrates an emotion control directive for training a speech-based LLM. Please follow these requirements:

\begin{enumerate}[leftmargin=1.5em]
    \item Retain the natural, conversational tone of the original question text.
    \item Convert all numerals into their English word equivalents.
    \item Generate an emotion control instruction based on the provided emotion: \textbf{\{emotion\}}. To enhance diversity, consider using synonyms or alternative expressions for the emotion. For example, you might say: ``Please explain this in a \{emotion\} voice'', ``Respond with a \{emotion\} tone'', ``Offer your explanation with a \{emotion\} sentiment'', or ``Convey this in a \{emotion\} manner''. Additionally, you can replace the \{emotion\} with one of its synonyms to further vary the expression.
\end{enumerate}

[question\_text]: \{question\_text\}

\vspace{1em}

Please output the result in the following JSON format:
\begin{verbatim}

{{

    "instruction_text": 
        {{instruction_text}}

}}

\end{verbatim}
\end{tcolorbox}

\subsection{Speed Control}

\begin{tcolorbox}[colback=gray!10, colframe=black,width=\linewidth,
                 arc=1mm, auto outer arc, boxrule=0.5pt,
                 breakable]
\textbf{Instruction:} \\
Below is an instruction to transform a user-provided conversational question text into an instruction text that integrates a speed control directive for training a speech-based LLM. Please follow these requirements:

\begin{enumerate}[leftmargin=1.5em]
    \item Retain the natural, conversational tone of the original question text.
    \item Convert all numerals into their English word equivalents.
    \item Generate a speed control instruction based on the provided speed: \textbf{\{speed\}}. To enhance diversity, consider using synonyms or alternative expressions for the speed. For example:
    \begin{itemize}
        \item ``Please explain this in a \{speed\} pace''
        \item ``Respond with a \{speed\} speed''
        \item ``Offer your explanation with a \{speed\} tempo''
        \item ``Convey this in a \{speed\} manner''
    \end{itemize}
\end{enumerate}

[question\_text]: \{question\_text\}

\vspace{1em}

Please output the result in the following JSON format:
\begin{verbatim}
{{

    "instruction_text": 
        {{instruction_text}}

}}
\end{verbatim}
\end{tcolorbox}

\subsection{Volume Control}

\begin{tcolorbox}[colback=gray!10, colframe=black,width=\linewidth,
                 arc=1mm, auto outer arc, boxrule=0.5pt,
                 breakable]
\textbf{Instruction:} \\
Below is an instruction to transform a user-provided conversational question text into an instruction text that integrates a volume control directive for training a speech-based LLM. Please follow these requirements:

\begin{enumerate}[leftmargin=1.5em]
    \item Retain the natural, conversational tone of the original question text.
    \item Convert all numerals into their English word equivalents.
    \item Generate a volume control instruction based on the provided volume: \textbf{\{volume\}}. To enhance diversity, consider using synonyms or alternative expressions for the volume. For example:
    \begin{itemize}
        \item ``Please explain this in a \{volume\} volume''
        \item ``Respond with a \{volume\} loudness''
        \item ``Offer your explanation with a \{volume\} intensity''
        \item ``Convey this in a \{volume\} manner''
    \end{itemize}
\end{enumerate}

[question\_text]: \{question\_text\}

\vspace{1em}

Please output the result in the following JSON format:
\begin{verbatim}
{{

    "instruction_text": 
        {{instruction_text}}

}}
\end{verbatim}
\end{tcolorbox}

\subsection{Accent Control}

\begin{tcolorbox}[colback=gray!10, colframe=black,width=\linewidth,
                 arc=1mm, auto outer arc, boxrule=0.5pt,
                 breakable]
\textbf{Instruction:} \\
Below is an instruction to transform a user-provided conversational question text into an instruction text that integrates an accent control directive for training a speech-based LLM. Please follow these requirements:

\begin{enumerate}[leftmargin=1.5em]
    \item Retain the natural, conversational tone of the original question text.
    \item Convert all numerals into their English word equivalents.
    \item Generate an accent control instruction based on the provided accent: \textbf{\{accent\}}. To enhance diversity, consider using synonyms or alternative expressions for the accent. For example:
    \begin{itemize}
        \item ``Please explain this using a \{accent\} accent''
        \item ``Respond with a \{accent\} intonation''
        \item ``Offer your explanation with expressions typical of a \{accent\} accent''
        \item ``Convey this in a manner typical of \{accent\} accent speech''
    \end{itemize}
\end{enumerate}

[question\_text]: \{question\_text\}

\vspace{1em}

Please output the result in the following JSON format:
\begin{verbatim}
{{

    "instruction_text": 
        {{instruction_text}}

}}
\end{verbatim}
\end{tcolorbox}

\subsection{Language Control}

\begin{tcolorbox}[colback=gray!10, colframe=black,width=\linewidth,
                 arc=1mm, auto outer arc, boxrule=0.5pt,
                 breakable]
\textbf{Instruction:} \\
Below is an instruction data for transforming a user-provided conversational text into an instruction text that integrates a specific language control directive for training a speech-based LLM. Please follow these requirements:

\begin{enumerate}[leftmargin=1.5em]
    \item Retain the natural, conversational tone of the original text.
    \item Convert all numerals into their word equivalents (in Chinese or English).
    \item Generate a language control instruction based on the provided language: \textbf{\{language\}}. To enrich the expression, consider using common expressions, idioms, or tones characteristic of that language. For example:
    \begin{itemize}
        \item ``Please explain this in \{language\}''
        \item ``Respond in \{language\}''
    \end{itemize}
    \item Generate the instruction\_text in English.
\end{enumerate}

[question\_text]: \{question\_text\}

\vspace{1em}

Please output the result in the following JSON format:
\begin{verbatim}
{{

    "instruction_text": 
        {{instruction_text}}

}}
\end{verbatim}
\end{tcolorbox}

\subsection{Composite Control}

\begin{tcolorbox}[colback=gray!10, colframe=black,width=\linewidth,
                 arc=1mm, auto outer arc, boxrule=0.5pt,
                 breakable]
\textbf{Instruction:} \\
Below is an instruction data for transforming a user-provided conversational text into an instruction text that integrates a natural, expressive style directive based on specific voice control attributes for training a speech-based LLM. Please follow these requirements:

\begin{enumerate}[leftmargin=1.5em]
    \item Extract only the \textbf{emotion}, \textbf{speech speed}, and \textbf{volume} from the provided description.
    \item Based on these three attributes, generate a concise, expressive \textbf{control style} phrase (within ten words). 
    \begin{itemize}
        \item This phrase should NOT list the attributes directly.
        \item Instead, describe the \textbf{feeling, delivery, or tone} implied by the combination.
        \item Examples:
        \begin{itemize}
            \item ``Barely contained rage spilling through sharp speech''
            \item ``Soft warmth with slow, deliberate rhythm''
            \item ``Urgent energy rising in a loud, fast tone''
        \end{itemize}
    \end{itemize}
    \item Retain the natural, conversational tone of the original instruction.
    \item Embed the generated control style naturally into the rewritten \texttt{instruction\_text}, without explicitly stating emotion/speed/volume.
\end{enumerate}

[question\_text]: \{question\_text\}

[description]: \{description\}

\vspace{1em}

Please output the result in the following JSON format:
\begin{verbatim}
{{
    "instruction_text": 
        {{instruction_text}},
    "style_description": 
        {{style_description}}
}}
\end{verbatim}
\end{tcolorbox}

\subsection{TTS Rendering Prompts}

\begin{tcolorbox}[colback=gray!10, colframe=black,width=\linewidth,
                 arc=1mm, auto outer arc, boxrule=0.5pt,
                 breakable]
\textbf{Emotion-based TTS Rendering Prompt:}\\
Please read the following text with the emotion of \{emotion\}: ``\{response\}''

\vspace{0.5em}
\textbf{Speed-based TTS Rendering Prompt:}\\
Please read the following text at a \{speed\} speaking rate: ``\{response\}''

\vspace{0.5em}
\textbf{Volume-based TTS Rendering Prompt:}\\
Please read the following text at a \{volume\} volume: ``\{response\}''

\vspace{0.5em}
\textbf{Composite Style TTS Rendering Prompt:}\\
Please read the text in the [Content] using the speaking style specified in the [Description]:

[Description]: \{description\}

[Content]: \{content\}

\end{tcolorbox}

\subsection{MOS Evaluation Prompt}\label{app:mos_eval}

\begin{tcolorbox}[colback=gray!10, colframe=black,width=\linewidth,
                 arc=1mm, auto outer arc, boxrule=0.5pt,
                 breakable]
\textbf{Instruction:} \\
Your task is to evaluate the alignment between the provided audio (a model's speech reply as a .wav file) and the given instruction. The instruction typically includes the content to be spoken (e.g., a question) and various style controls (e.g., emotion, speed, volume, accent, style description, language change).

The evaluation should focus on how well the audio reply adheres to \textbf{ALL} aspects of the provided instruction. This includes, but is not limited to:
\begin{itemize}
  \item Accuracy of the spoken content based on the ``question'' or core message in the instruction.
  \item Consistency with the specified \texttt{emotion}, if any.
  \item Adherence to the specified \texttt{speed} (e.g., fast, slow, normal), if any.
  \item Adherence to the specified \texttt{volume} (e.g., loud, soft, normal), if any.
  \item Correctness and consistency of the specified \texttt{accent} (e.g., British English, American English, specific regional accent), if any.
  \item Realization of the \texttt{style description} (e.g., ``energetic and friendly,'' ``formal and serious''), if any.
  \item Correct execution of any \texttt{language change} instruction (e.g., ``switch to French for the last sentence''), if any.
\end{itemize}

Rate the audio on a scale of 1 to 5 based on the following criteria:

\begin{itemize}
  \item \textbf{1 point:} The audio completely fails to follow the instruction. The spoken content may be incorrect, significantly incomplete, or unintelligible. Multiple specified style controls are ignored, misapplied, or contradictory to the instruction.
  \item \textbf{2 points:} The audio attempts to follow parts of the instruction but does so poorly or inconsistently. There may be major deviations in spoken content, or several style controls are noticeably incorrect, faint, mismatched, or missing. The overall result significantly deviates from the instruction.
  \item \textbf{3 points:} The audio generally follows the main aspects of the instruction (e.g., content is mostly accurate, dominant style controls are attempted). However, there are noticeable inconsistencies in one or more style controls, some secondary style controls are not adequately met, or the overall delivery has clear flaws in matching the full instruction.
  \item \textbf{4 points:} The audio effectively follows most aspects of the instruction with only minor imperfections or slight deviations in one or two style control elements. The spoken content is accurate, and the overall delivery strongly aligns with the instruction. Most specified style controls are well-realized.
  \item \textbf{5 points:} The audio perfectly matches all aspects of the provided instruction. The spoken content is accurate. All specified style controls (emotion, speed, volume, accent, style description, language change, etc., as applicable) are flawlessly, clearly, and appropriately executed, resulting in a natural and fully compliant delivery.
\end{itemize}

Please note: Your evaluation should be independent and strictly based on the provided instruction and the audio's alignment with it. Consider all specified parameters in the instruction.\\

Please provide your score based on the audio's adherence to \textbf{ALL} specified elements in the instruction.

Below is the transcription of user's instruction:\\
\textbf{[Instruction Details]}\\
\{instruction\}

After evaluating, please output \textbf{ONLY} the final calculated score (a number between 1.0 and 5.0, rounded to the nearest 0.5) without anything else.

Please strictly follow the standards and avoid leniency in your evaluation. Ensure that the score reflects the exact alignment between the audio and the full instruction, without overestimating or underestimating the quality.
\end{tcolorbox}

\subsection{IFR Evaluation Prompt}\label{app:ifr_eval}

\begin{tcolorbox}[colback=gray!10, colframe=black,width=\linewidth,
                 arc=1mm, auto outer arc, boxrule=0.5pt,
                 breakable]
\textbf{Instruction:} \\
Your task is to determine whether the provided audio strictly follows the acoustic control instructions in the given prompt. Focus \textbf{ONLY} on the acoustic aspects and output \textbf{ONLY 1} (follows instruction) or \textbf{0} (does not follow instruction).

The evaluation should check for:

\begin{enumerate}[leftmargin=1.5em]
  \item \textbf{Emotion Control (if specified):} Does the voice express the exact requested emotion?
  \item \textbf{Speed Control (if specified):} Does the speech maintain the exact requested pace?
  \item \textbf{Volume Control (if specified):} Does the audio maintain the exact requested volume level?
  \item \textbf{Accent Control (if specified):} Does the voice use the exact requested accent?
  \item \textbf{Style Description (if specified):} Does the voice match the exact style descriptors?
  \item \textbf{Language Switch (if specified):} Does the voice switch to the requested language at the specified point?
\end{enumerate}

\textbf{IMPORTANT:}
\begin{itemize}
  \item Output ONLY 1 or 0
  \item 1 = ALL specified controls are correctly followed
  \item 0 = ANY specified control is not followed correctly
  \item Ignore content accuracy completely
  \item Consider ONLY the specified control elements
\end{itemize}

Below is the instruction with acoustic controls:\\
\textbf{[Instruction Details]}\\
\{instruction\}

After evaluating, output \textbf{ONLY 1 or 0}.
\end{tcolorbox}

\end{document}